\documentclass[%
 aip,jcp,
 amsmath,amssymb,
 reprint,
nolongbibliography
]{revtex4-2}

\usepackage[a-1b]{pdfx} 

\usepackage{graphicx}% Include figure files
\usepackage{dcolumn}% Align table columns on decimal point
\usepackage{bm}% bold math
\usepackage[utf8]{inputenc}
\usepackage[T1]{fontenc}
\usepackage{mathptmx}
\usepackage{listings}
\usepackage{rotating} % Rotating table

\usepackage{color}
\usepackage{dcolumn} % decimal align in tables
\usepackage{bm} % bold math
\usepackage{graphicx}
\usepackage{multirow} % for table cells to span rows
\usepackage{pifont} % for checkmarks
\usepackage{epsfig}
\usepackage{amsmath} % matrix
\usepackage{amssymb}
\usepackage{mathrsfs}
\usepackage{subfigure}
\usepackage{float}
\usepackage{booktabs}
\usepackage{tabularx}
\usepackage{natbib}
\usepackage{gensymb}
\usepackage{rotating}
\usepackage{threeparttable}
\usepackage{comment}
\usepackage{physics} % lazy bra-ket
\usepackage[normalem]{ulem}
\usepackage{nicefrac}
\hypersetup{
  colorlinks,
  linkcolor=blue,
  citecolor=blue,
  urlcolor=blue
} 

\begin{document}

\author{Stephen H. Yuwono }
\affiliation{
             Department of Chemistry and Biochemistry,
             Florida State University,
             Tallahassee, FL 32306-4390}   
             
\author{Run R. Li}
\affiliation{
             Department of Chemistry and Biochemistry,
             Florida State University,
             Tallahassee, FL 32306-4390}    

\author{Tianyuan Zhang}
\affiliation{Department of Chemistry, University of Washington, Seattle, WA 98195, USA}

\author{Xiaosong Li}
\affiliation{Department of Chemistry, University of Washington, Seattle, WA 98195, USA}
             
\author{A. Eugene DePrince III}
\email{adeprince@fsu.edu}
\affiliation{
             Department of Chemistry and Biochemistry,
             Florida State University,
             Tallahassee, FL 32306-4390}

\title{Two-component relativistic equation-of-motion coupled cluster for electron ionization}

\begin{abstract}
We present an implementation of relativistic ionization-potential (IP) equation-of-motion coupled-cluster (EOMCC) with up to 3-hole--2-particle (3h2p) excitations that makes use of the molecular mean-field exact two-component (mmfX2C) framework and the full Dirac--Coulomb--Breit Hamiltonian. The closed-shell nature of the reference state in an X2C-IP-EOMCC calculation allows for accurate predictions of spin-orbit splittings in open-shell molecules without breaking degeneracies, as would occur in an excitation-energy EOMCC calculation carried out directly on an unrestricted open-shell reference. We apply X2C-IP-EOMCC to the ground and first excited state of the HCCX$^+$ (X = Cl, Br, I) cations, where it is demonstrated that a large basis set (\emph{i.e.}, quadruple-zeta quality) and 3h2p correlation effects are necessary for accurate absolute energetics. The maximum error in calculated adiabatic IPs is on the order of 0.1 eV, whereas spin-orbit splittings themselves are accurate to $\approx 0.01$ eV, as compared to experimentally obtained values.
\end{abstract}

\maketitle

\section{Introduction}

\label{SEC:INTRODUCTION}

Open-shell molecular systems arise in a variety of interesting contexts, including radical intermediate species in chemical reactions,\cite{Curran16_58,Waser22_7344,Stahl22_405} spin-crossover complexes,\cite{Nicolazzi11_3313,Shatruk21_14563,Dowben22_1742} single-molecule magnets,\cite{Ruiz20_9916,Coronado20_87,Chilton22_79,Johansson24_031305} and molecular spin qubits.\cite{Coronado19_301,Alexandrova21_9567,Sinitskii23_3531} An accurate {\em ab initio} description of such systems is critical and yet can be challenging due to the significant multi-reference (MR) character that open-shell molecules often display, as opposed to the mostly single-reference (SR) nature of closed-shell systems. Rather than deal with the complexities and subtleties of MR electronic structure theory, which can be non-trivial, it is possible to treat open-shell molecular systems in a simple, reliable way using SR methodologies, given a suitably chosen {\em ansatz}.

A popular SR framework for many-body electronic structure calculations is coupled-cluster (CC) theory, \cite{Coester58_421,Kuemmel60_477,Cizek66_4256,Cizek69_35,Shavitt72_50,Li99_1,Musial07_291} which can be extended to the description of excited states via the equation-of-motion (EOM)\cite{Emrich81_379,Bartlett89_57,Bartlett93_7029} formalism. CC/EOMCC theory forms a family of robust, reliable, and systematically improvable approaches for the high-accuracy description of correlated many-electron systems. Particularly accurate results can be obtained when one considers methods that go beyond the basic CC/EOMCC with single and double excitations (CCSD/EOMCCSD),\cite{Bartlett82_1910,Zerner82_4088,Bartlett93_7029} such as those including up to triple (CCSDT/EOMCCSDT)\cite{Schaefer86_207,Bartlett87_7041,Schaefer88_382,Bartlett90_6104} or quadruple excitations (CCSDTQ/EOMCCSDTQ),\cite{Adamowicz91_6645,Bartlett91_387,Bartlett92_4282,Adamowicz94_5792} and this hierarchy quickly converges to the full configuration interaction (CI) limit.\cite{Musial07_291} Despite these strengths, the direct application of CC/EOMCC to open-shell systems can be problematic. For example, spin contamination effects associated with an unrestricted or generalized Hartree--Fock (UHF or GHF, respectively) reference configuration can lead to a loss in proper spin-degeneracy structure, which can be a significant issue in some applications. 

An alternative strategy that still falls within the CC/EOMCC family is to begin with a reference state that is well-described by a SR method but differs from the target open-shell state by particle number. For example, the ionization potential (IP) EOMCC approach\cite{Stanton94_65,Snijders92_55,Snijders93_15,Gauss94_8938,Bartlett03_1128,Bartlett04_210,Gauss05_154107,Wloch05_134113,Wloch06_2854,Piecuch06_234107} begins with a closed-shell $N$-electron state, which should be well-described by the CC hierarchy applied to a restricted Hartree--Fock (RHF) reference configuration, and parametrizes the open-shell $(N-1)$-electron state (the ionized state) using non-particle-conserving excitation operators that remove an electron from the reference state. The key benefit of this approach is that, by starting from properly spin-adapted reference wave function that does not break desired orbital degeneracies, the subsequent open-shell energetics should retain this property.  This advantage becomes more apparent when considering energy splittings due to spin--orbit coupling (SOC) effects in relativistic calculations, where spin-contamination-induced degeneracy issues could lead to energy errors on the order of the splittings themselves. Moreover, even though $S_z$ and $S^2$ are technically not good quantum numbers upon the introduction of SOC, one can still take advantage of time-reversal symmetry in the form of Kramers pairs for closed-shell determinants, which in practice produces degenerate pairs of spinors as one would obtain in the non-relativistic RHF case.\cite{Kramers30_959,Messiah_book,Li20_090903} This symmetry preservation extends the benefits of IP-EOMCC to the relativistic domain, as the underlying structure of relativistic CC calculations would be analogous to that of non-relativistic singlet spin-adapted CC.\cite{Lee95_411}

This work uses CC/IP-EOMCC to examine ionization potentials of HCCX (X = Cl, Br, I) systems, which have been characterized both experimentally and theoretically, from as far back as the 1970s.\cite{Kloster-Jensen70_1073,Maier77_1406,Ochsner85_3181,Ochsner85_1587,Maier88_45,Dreizler89_296,Boggs03_159,Merkt13_9353,Yang22_857348,Li22_866137} More recently, the theoretical study presented in Ref.~\citenum{Schlegel24_117313} provided great details about the geometries, vibrational frequencies, and both vertical and adiabatic IP values of these molecules. However, that study did not account for relativistic effects, which are necessary to describe SOC-induced splitting of the doublet ground and excited states of HCCX$^+$.  Here, we apply relativistic CC/IP-EOMCC to this problem, and the relativistic treatment we employ falls within the exact two-component (X2C)
\cite{Dyall97_9618,Dyall98_4201,Enevoldsen99_10000,Dyall01_9136,Cremer02_259,Liu05_241102,Peng06_044102,Cheng07_104106,Saue07_064102,Peng09_031104,Liu10_1679,Liu12_154114,Reiher13_184105,Li16_3711,Li16_104107,Repisky16_5823,Li17_2591,Cheng21_e1536,Li22_2947,Li22_2983,Li22_5011,Li24_3408,Li24_7694,Li24_041404} family of methods. 
There are a number of prior studies that extend the IP-EOMCC approach to the relativistic domain using X2C\cite{Pal15_115009,Liu17_3713,SeveroPereiraGomes18_174113,Cheng19_1642,SeveroPereiraGomes21_3583} or other\cite{Fan07_024104,Li12_174102,Piecuch14_101102,Pal14_042510,Pal14_062501,Krylov15_064102,Piecuch16_084306,Pal16_074110,Nakajima17_827,Chattopadhyay18_042705,Wang18_935} relativistic frameworks. Focusing on the X2C-based calculations, the majority of these investigations used spin-free X2C, and the most sophisticated calculations have considered only the Dirac--Coulomb (DC) or Dirac--Coloumb--Gaunt (DCG) Hamiltonians, as opposed to the full Dirac--Coulomb--Breit (DCB) Hamiltonian.\cite{SeveroPereiraGomes18_174113,SeveroPereiraGomes21_3583} As for correlation effects, to the best of our knowledge, only one of these studies has gone beyond the IP-EOMCCSD level of theory; Ref.~\citenum{Cheng19_1642} treated K-edge ionization features at up to the IP-EOMCCSDTQ level, but it should be noted that these high-order correlation effects were assessed only within a spin-free X2C formalism. The present study aims to assess high-order correlation effects (up to triple excitations in both the ground-state CC and IP-EOM parts of the algorithm), as well as a mean-field treatment of the full DCB Hamiltonian through the molecular mean-field (mmf)\cite{Ilias09_124116,Visscher14_041107,SeveroPereiraGomes18_174113,Li24_3408} X2C formalism.

The remainder of this paper is organized in the following manner. Section \ref{SEC:THEORY} provides the pertinent details of the X2C and CC/IP-EOMCC approaches. Section \ref{SEC:COMPUTATIONAL_DETAILS} describes the details of our computations, the results of which are presented in Section \ref{SEC:RESULTS}, including analyses of geometry, basis set, and correlation effects. Some concluding remarks can be found in Section \ref{SEC:CONCLUSIONS}.

\section{Theory}
\label{SEC:THEORY}

\subsection{The X2C relativistic framework}
\label{SUB:X2C}

The relativistic frameworks used in this work are the mmf formulation of X2C and a more approximate one-electron X2C approach (1eX2C). Detailed comparisons of mmfX2C and 1eX2C can be found in Refs.~\citenum{Li24_3408} and \citenum{Ilias09_124116} and the references contained therein. In this Section, we only summarize the details that are relevant to this work. 

The X2C approach seeks to incorporate  relativistic effects into the electronic Hamiltonian, the second-quantized form of which is
\begin{equation}
\label{eqn:electronic_hamiltonian}
\hat{H} = h_\mu^\nu \hat{a}^{\mu} \hat{a}_{\nu} +
\frac{1}{2}g^{\lambda\sigma}_{\mu\nu}\hat{a}^{\mu} \hat{a}^{\nu} \hat{a}_{\sigma} \hat{a}_{\lambda}
\end{equation}
The symbol $h_\mu^\nu$ represents an element of the core Hamiltonian matrix, 
    $g_{\mu \nu}^{\lambda \sigma} = 
    \braket*{\mu \nu}{\lambda \sigma}$
is a two-electron integral in physicists' notation, and $\hat{a}_{\mu}$ ($\hat{a}^{\mu}$) is the fermionic annihilation (creation) operator associated with the atomic basis function $\mu$. Here and throughout the remainder of this work, we employ the Einstein summation convention where repeated upper and lower indices imply a sum.

To understand how the X2C transformation is done, let us consider the 4c Dirac equation and the matrix representation of the Dirac Hamiltonian, which, with the restricted kinetic-balance prescription, has the form 
\begin{align}
\label{eqn:dirac_hamiltonian}
\mathbf{H}^{4\rm c}_{\rm D} = 
    \begin{pmatrix}
        \mathbf{V} & \mathbf{T} \\
        \mathbf{T} & \frac{1}{4c^2}\mathbf{W}-\mathbf{T}
    \end{pmatrix}
\end{align}
Here, $\mathbf{V}$ is the 2c matrix representation of the scalar potential operator, $\mathbf{T}$ is the 2c matrix representation kinetic energy operator, $\mathbf{W}$ is the 2c relativistic potential matrix, and $c$ is the speed of light. The relativistic potential matrix is
\begin{align}
\mathbf{W} = (\bm{\sigma}\cdot\mathbf{p}) \mathbf{V} (\bm{\sigma}\cdot\mathbf{p})\label{eq:W}
\end{align}
where $\bm{\sigma}$ and $\mathbf{p}$ are vectors of the Pauli matrices and the matrix representation of the linear momentum operator, respectively. The Dirac equation is
\begin{align}
    \begin{pmatrix}
        \mathbf{V} & \mathbf{T} \\
        \mathbf{T} & \frac{1}{4c^2}\mathbf{W}-\mathbf{T}
    \end{pmatrix}
    &
    \begin{pmatrix}
        \mathbf{C}_\mathrm{L}^+ & \mathbf{C}_\mathrm{L}^- \\
        \mathbf{C}_\mathrm{S}^+ & \mathbf{C}_\mathrm{S}^-
    \end{pmatrix}
     = \nonumber \\
     &\begin{pmatrix}
        \mathbf{S} & \mathbf{0} \\
        \mathbf{0} & \frac{1}{2c^2}\mathbf{T}
    \end{pmatrix}
    \begin{pmatrix}
        \mathbf{C}_\mathrm{L}^+ & \mathbf{C}_\mathrm{L}^- \\
        \mathbf{C}_\mathrm{S}^+ & \mathbf{C}_\mathrm{S}^-
    \end{pmatrix}
    \begin{pmatrix}
        \bm{\epsilon}^+ & \mathbf{0} \\
        \mathbf{0} & \bm{\epsilon}^-
    \end{pmatrix} \label{eq:Dirac4C}
\end{align}
where $\mathbf{S}$ is the overlap matrix, $\mathbf{C}_{\rm L}$ and $\mathbf{C}_{\rm S}$ represent the large- and small-components of the molecular orbital coefficients respectively, the $\bm{\epsilon}$ matrices are diagonal matrices with the energy eigenvalues on the diagonal, and the superscripts $\pm$ refer to positive- and negative-energy solutions. For quantum chemistry applications, we are only interested in the positive-energy solutions, so it is convenient to define the X2C transformation as the unitary transformation that block diagonalizes the Dirac Hamiltonian as
\begin{equation}
\label{eqn:x2c_unitary}
    \mathbf{U}^\dagger \mathbf{H}_{\rm D}^\mathrm{4c} \mathbf{U} = 
    \begin{pmatrix}
        \mathbf{H}^+  & \mathbf{0} \\
        \mathbf{0} & \mathbf{H}^-
    \end{pmatrix}
\end{equation}
while also satisfying the renormalization condition that the same overlap matrix will be used when diagonalizing $\mathbf{H}^+$ as would be used when diagonalizing a non-relativistic one-body Hamiltonian matrix.\cite{Reiher13_184105}
Given this transformation, the eigenvalues and eigenvectors of the Hamiltonian $\mathbf{H}^+$ can be determined in a 2c rather than 4c space. 

In the 1eX2C approach, ${\bf V}$ in Eqs.~(\ref{eqn:dirac_hamiltonian}) and (\ref{eq:W}) is the nuclear potential that gives rise to scalar and spin--orbit relativistic effects, without any two-body contributions from the Coulomb or Breit operators. As a result, the two-electron part of the Hamiltonian operator does not affect the 4c to 2c transformation, and the transformation matrix can be constructed in a single step. 
%In the 1eX2C approach, relativistic effects are incorporated in the electronic core Hamiltonian through the the matrix elements $h_\mu^\nu$ that enter Eq.~(\ref{eqn:electronic_hamiltonian}), resulting in a one-step procedure to construct the transformation matrix. 
%Note that the two-electron part of the Hamiltonian operator does not affect the 4c to 2c 1eX2C transformation. 
In order to account for missing two-body spin--orbit interaction effects in 1eX2C, 
the spin--orbit part of $h_\mu^\nu$
is scaled using a screened nuclear spin--orbit (SNSO) factor; 
in this work, we use the row-dependent DCB-parameterized factors of Ref.~\citenum{Li23_5785}.  The 1eX2C Hartree--Fock (HF) procedure is carried out in the 2c space with this modified Hamiltonian with an untransformed (or bare) two-electron Coulomb operator, and the post-HF portion of the calculation then proceeds as usual in the 2c space, using the same Hamiltonian.

In the mmfX2C approach, two-body effects are included in the molecular potential via the Dirac--Coulomb (DC)/Dirac--Coulomb--Breit operators, in which case Eq.~(\ref{eq:Dirac4C}) represents the 4c Dirac--Hartree--Fock (DHF) equation. This equation must be solved self-consistently, rather than via the one-step diagonalization in 1eX2C. As a result, two-electron relativistic effects are fully accounted for at the mean-field level.
%In the mmfX2C approach, the two-electron DC/DCB operator is included in the 4c Dirac--Hartree--Fock (DHF) equation (Eq.~\ref{eq:Dirac4C}). As a result, Equation \ref{eq:Dirac4C} must be solved self-consistently, instead of the one-step diagonalization in 1eX2C, to obtain the eigenvalues and eigenvectors. In this approach, two-electron relativistic effects are fully accounted for at the mean-field level.
%the reference configuration in a 1eX2C-based post-self-consistent-field (SCF) method is obtained from a 2c HF calculation carried out with the SNSO-scaled Hamiltonian, as described above.  On the other hand, the reference configuration in the mmfX2C scheme is  a 2c function obtained from the 4c$\rightarrow$2c transformation of a 4c Dirac--Hartree--Fock (DHF) wave function. 
In this work, the DHF procedure makes use of the full Coulomb--Breit operator to describe the two-body part of the Hamiltonian,\cite{Li21_3388,Li22_064112,Li23_171101}
\begin{equation}
\label{eqn:breit_hamiltonian}
    V(r_{ij}) = \frac{1}{r_{ij}} - \frac{1}{2} \left( \frac{\bm{\alpha}_i\cdot\bm{\alpha}_j}{r_{ij}} + \frac{\bm{\alpha}_i\cdot\mathbf{r}_{ij}\bm{\alpha}_j\cdot\mathbf{r}_{ij}}{r_{ij}^3} \right)
\end{equation}
where the $\bm{\alpha}_i$ are vector quantities with
\begin{equation}
\label{eqn:alpha_matrix}
    \bm{\alpha}_{i,q} =
    \begin{pmatrix}
        \mathbf{0_2} & \bm{\sigma}_q \\
        \bm{\sigma}_q & \mathbf{0_2}
    \end{pmatrix},\;q=\{x,y,z\}
\end{equation}
Here, $\bm{\sigma}_x$, $\bm{\sigma}_y$, and $\bm{\sigma}_z$ represent the Pauli spin matrices. After the 4c DHF self-consistent-field procedure, the 2c reference function and corresponding 2c Fock matrix are obtained from the X2C transformation. The subsequent correlated calculation then makes use of the X2C-transformed Fock matrix and two-electron integrals evaluated using an untransformed Coulomb operator, within the 2c space.
For the remainder of this work, we denote mmfX2C carried out with the DCB Hamiltonian as DCB-X2C.

In terms of the implementation, we recall that the 4c expansion of any atomic-orbital integrals should be done in a fully uncontracted basis to prevent variational collapse or prolapse of the 4c DHF solution.\cite{FaegriJr01_252,Mochizuki03_399,Mochizuki03_40,Li21_207,Haiduke22_1901,Li21_3388,Li22_064112,Li23_171101} However, after performing the X2C transformation, one can recontract the 2c integrals using the basis set's original contraction coefficients.\cite{Li24_3408} 
Given that we are working in a 2c space only, the post self-consistent field (SCF) calculations also employ the no-virtual-pair approximation,\cite{Mittleman71_893,Sucher80_348,Sucher84_703,Hess86_3742,Peierls97_552} in which electron--positron correlation effects are ignored.

\subsection{IP-EOMCC theory}
\label{SUB:IPEOM}

This section provides an overview of the key details of CC theory and its extension to the description of ionized states via IP-EOMCC. All quantities are represented within the molecular spinor basis determined via the underlying 1eX2C- or mmfX2C-HF procedure. The labels $i_n$ and $a_n$ (with $n = 1, 2, 3, ...$) refer to molecular spinors that are occupied or unoccupied in the reference configuration. 

The ground-state CC wave function for an $N$-electron state is parametrized as
\begin{align}
    \ket*{\Psi_0^{(N)}} = \exp(\hat{T})\ket{\Phi_0},
\end{align}
where  $|\Phi_0\rangle$ is a 2c reference determinant, and $\hat{T}$ is the cluster operator, which can be expanded in terms of $n$-particle--$n$-hole ($n$p$n$h) excitation components as
\begin{equation}
\label{eqn:ccsd_t_operator}
    \hat{T} = \sum_{n=1}^{M} \hat{T}_n,\;
    \hat{T}_n = \left(\frac{1}{n!}\right)^2 t_{a_1 \ldots a_n}^{i_1 \ldots i_n} \prod_{k=1}^{n} (\hat{a}^{a_k} \hat{a}_{i_k}).
\end{equation}
Here, $M$ is a pre-selected truncation level, and $t_{a_1 \ldots a_n}^{i_1 \ldots i_n}$ are $n$-body cluster amplitudes. 
The choice of $M$ gives the usual CCSD ($M=2$), CCSDT ($M=3$), CCSDTQ ($M=4$), \emph{etc.}~hierarchy of SRCC methods, and when $M=N$, one reaches the full CC limit, which provides a description of the system that is equivalent to that which would be obtained from full CI.
The cluster amplitudes are obtained by solving the projective equations
\begin{equation}
\label{eqn:ccsd_projective}
    \mel*{\Phi_{i_1 \ldots i_n}^{a_1 \ldots a_n}}{\bar{H}}{\Phi} = 0 \; \forall \; \ket*{\Phi_{i_1 \ldots i_n}^{a_1 \ldots a_n}}, n = 1, 2, ..., M,
\end{equation}
where $\bar{H} = e^{-\hat{T}} \hat{H} e^{\hat{T}}$ is the similarity-transformed Hamiltonian, and $\ket*{\Phi_{i_1 \ldots i_n}^{a_1 \ldots a_n}}$ represents a Slater determinant that is $n$-tuply substituted relative to $|\Phi_0\rangle$.  Given optimal amplitudes, the ground-state CC energy is then obtained by computing the expectation value
\begin{equation}
\label{eqn:ccsd_energy}
    E_0 = \mel*{\Phi_0}{\bar{H}}{\Phi_0}.
\end{equation}

Energies and wave functions of the $(N-1)$-electron ({\em i.e.}, ionized) states can be obtained from IP-EOMCC, where the $K$-th ionized state ($K>0$) is parametrized as 
\begin{equation}
\label{eqn:eom_wfn}
    \ket*{\Psi_K^{(N-1)}} = \hat{R}_K \ket*{\Psi_0^{(N)}} = \hat{R}_K \exp(\hat{T}) \ket{\Phi_0},
\end{equation}
Here, $\hat{R}_K$ is a linear, non-particle-conserving operator that removes one electron from the $N$-electron state. In a similar manner as was done for the cluster operator, $\hat{R}_K$ can be expanded by excitation order as
\begin{equation}
\label{eqn:eom_r_operator}
    \hat{R}_K = \sum_{n=1}^{M^\prime} \hat{R}_{K,n},\;
    \hat{R}_{K,n} = \frac{1}{n!(n-1)!} r_{K,a_2 \ldots a_{n}}^{\phantom{K,}i_1 \ldots i_n}
                     \hat{a}_{i_1} \prod_{k=2}^{n} (\hat{a}^{a_k} \hat{a}_{i_k}),
\end{equation}
where $r_{K,a_2 \ldots a_n}^{\phantom{K,}i_1 \ldots i_n}$ is an excitation amplitude and $M^\prime$ is a pre-selected truncation level that is not necessarily equal to $M$. Inserting Eq.~(\ref{eqn:eom_wfn}) into the time-independent Schr{\"o}dinger equation leads to the IP-EOMCC eigenvalue equation
\begin{equation}
\label{eqn:eom_eigenvalue}
    \bar{H} \hat{R}_K \ket{\Phi_0} = E_K \hat{R}_K \ket{\Phi_0}
\end{equation}
or
\begin{equation}
\label{eqn:eom_eigenvalue_commutator}
    [\bar{H},\hat{R}_K] \ket{\Phi_0} = \omega_K \hat{R}_K \ket{\Phi_0},
\end{equation}
where $\omega_K = E_K - E_0$ is the vertical IP of the $K$-th $(N-1$)-electron state. 

By setting $M^\prime=M$ in Eq.~(\ref{eqn:eom_r_operator}), one arrives at the usual IP-EOMCCSD ($M^\prime=M=2$), IP-EOMCCSDT ($M^\prime=M=3$), \emph{etc.}~methods, but, again, $M^\prime$ and $M$ need not be equal in general. As an example, in the IP-EOMCCSD(3h2p)\cite{Wloch05_134113,Wloch06_2854} approach, one diagonalizes the CCSD ($M=2$) similarity-transformed Hamiltonian in the space spanned by $\ket*{\Phi_i}$, $\ket*{\Phi_{ij}^{\phantom{i}b}}$, and $\ket*{\Phi_{ijk}^{\phantom{i}bc}}$ determinants ($M^\prime=3$), which refer to configurations that are generated from $|\Phi_0\rangle$ by the removal of an electron, plus up to two-particle transitions. It has been shown\cite{Wloch05_134113,Wloch06_2854} that the correlation effects folded into the CCSD $\bar{H}$, \emph{i.e.}, excluding $\hat{T}_3$, may be sufficient for IP-EOMCCSD(3h2p) to describe the multi-reference character of open-shell radicals,
even though the 3h2p transitions correspond to parts of 
triple excitations in the ($N-1$)-electron manifold. This mismatch in excitation level may overcorrelate the ionized states relative to the $N$-electron ground state, leading to absolute IP values that are too low. However, the electronic spectrum of the $(N-1)$-electron species relative to the $K=1$ state may be improved compared to that determined via IP-EOMCCSD.\cite{Wloch05_134113,Wloch06_2854} IP-EOMCCSD(3h2p) is also less computationally demanding than full IP-EOMCCSDT due to the neglect of $\hat{T}_3$ in the ground-state CC part of the algorithm and in $\bar{H}$. Additional details about the IP-EOMCCSD(3h2p) approach can be found in Refs.~\citenum{Wloch05_134113,Wloch06_2854,Piecuch06_234107}.

\section{Computational Details}
\label{SEC:COMPUTATIONAL_DETAILS}

This work investigates the accuracy of the relativistic IP-EOMCC approach applied to molecular systems. Specifically, we consider the  X $^2\Pi_\Omega$ and A $^2\Pi_\Omega$ states of the haloacetylene cations (HCCX$^+$, X = Cl, Br, I), for which the SOC splitting between the $\Omega = {\nicefrac{3}{2}}$ and ${\nicefrac{1}{2}}$ components in the photoelectron spectra have been experimentally resolved. \cite{Kloster-Jensen70_1073,Maier77_1406} 
The relativistic treatments are based on either the DCB-X2C formalism or the 1eX2C approximation with DCB-parameterized SNSO scaling factors.\cite{Li23_5785} 
In addition to IP-EOMCCSD, we also consider the IP-EOMCCSD(3h2p) and IP-EOMCCSDT approaches, which allow us to assess higher-order correlation effects not captured by IP-EOMCCSD.

Section \ref{SEC:RESULTS} begins with a study of the effect that the basis set has on 
vertical IPs to the X and A $^2\Pi_\Omega$ ($\Omega = {\nicefrac{3}{2}}$ and ${\nicefrac{1}{2}}$) states computed at the IP-EOMCCSD level of theory. For this study, structures for neutral HCCCl and HCCBr were taken from NIST's Computational Chemistry Comparison and Benchmark Database,\cite{CCCBDB}
whereas the structure for HCCI was taken from experimental data reported in Ref.~\citenum{Boggs03_159}. The structural parameters are reproduced in Table \ref{tab:hccx_geom}. 
The basis sets examined were the ANO-RCC-VDZP, ANO-RCC-VTZP, and ANO-RCC-VQZP basis sets,\cite{Widmark05_6575} the x2c-SVPall-2c, x2c-TZVPPall-2c, and x2c-QZVPPall-2c sets,\cite{Weigend17_3696,Weigend20_5658} and the Dyall v2z, v3z, and v4z sets.\cite{Dyall98_366,Dyall02_335,Dyall06_441,Dyall16_128,DyallBasisZenodo} 
We also used the cc-pV$n$Z-DK\cite{Dixon01_48} ($n = {}$D, T, Q, 5) basis sets for HCCCl and HCCBr to estimate the remaining basis-set effects beyond the quadruple-$\zeta$ level (see the Supplemental Information).
The ANO-RCC-, x2c-, and cc-type basis sets were taken from the Basis Set Exchange,\cite{Feller96_1571,Windus07_1045,Windus19_4814} while the uncontracted Dyall bases were obtained from Ref.~\citenum{DyallBasisZenodo}.

\begin{table}[!htbp]
    \centering
    \caption{Equilibrium bond lengths of the neutral HCCX molecules used in the basis set analysis for vertical IPs computed at the IP-EOMCCSD level of theory.}
    \label{tab:hccx_geom}
    \begin{tabular}{@{\extracolsep{20pt}} cccc}
        \hline \hline
        Bond & X = Cl\cite{CCCBDB} & X = Br\cite{CCCBDB} & X = I\cite{Boggs03_159} \\
        \hline
        X-C  & 1.6368 & 1.7916 & 1.9891 \\
        C-C  & 1.2033 & 1.2038 & 1.2058 \\
        C-H  & 1.0550 & 1.0552 & 1.0624 \\
        \hline \hline
    \end{tabular}
\end{table}

The accuracy of X2C-based IP-EOMCC was also evaluated against the adiabatic data for the haloacetylene molecules reported in Ref.~\citenum{Maier77_1406}. For this study, the geometries for each of the molecules are taken from Ref.~\citenum{Schlegel24_117313}, which were optimized using the cc-pVTZ\cite{Dunning89_1007,Dunning93_1358,Stoll06_13877,Dolg03_11113} basis set at the non-relativistic CCSD level of theory, in the case of the neutral species, and non-relativistic IP-EOMCCSD, in the case of the cation states. The resulting geometric parameters are reproduced in Table \ref{tab:hccx_geom_adiabatic}. In the adiabatic IP calculations, where we focus on the DCB-X2C relativistic framework, we employed IP-EOMCCSD using the ANO-RCC-VDZP and ANO-RCC-VQZP basis sets, as well as IP-EOMCCSD(3h2p) and IP-EOMCCSDT calculations in the ANO-RCC-VDZP basis.

\begin{table}[!htbp]
    \centering
    \caption{Equilibrium bond lengths of the HCCX molecules in their ground state and the first- and second-ionized states reported in Ref.~\citenum{Schlegel24_117313}.}
    \label{tab:hccx_geom_adiabatic}
    \begin{tabular}{@{\extracolsep{10pt}} ccccc}
        \hline \hline
        System & Bond & X = Cl & X = Br & X = I \\
        \hline
        \multirow{3}{*}{Neutral HCCX}
        & X-C  & 1.648 & 1.793 & 1.997 \\
        & C-C  & 1.202 & 1.203 & 1.206 \\
        & C-H  & 1.060 & 1.061 & 1.062 \\
        \hline
        \multirow{3}{*}{HCCX$^+$ (X $^2\Pi_\Omega$)}
        & X-C  & 1.577 & 1.725 & 1.934 \\
        & C-C  & 1.236 & 1.231 & 1.222 \\
        & C-H  & 1.074 & 1.073 & 1.071 \\
        \hline
        \multirow{3}{*}{HCCX$^+$ (A $^2\Pi_\Omega$)}
        & X-C  & 1.771 & 1.903 & 2.096 \\
        & C-C  & 1.210 & 1.218 & 1.233 \\
        & C-H  & 1.071 & 1.071 & 1.072 \\
        \hline \hline
    \end{tabular}
\end{table}

All calculations reported in this work were performed with the IP-EOMCC code implemented using the TiledArray tensor algebra framework\cite{TiledArray} in a development branch of Chronus Quantum.\cite{Li20_e1436} Initial prototyping of the IP-EOMCC code was done with the help of the \texttt{p$\dagger$q} automated code generator.\cite{DePrince21_e1954709} We have verified the correctness of non-relativistic IP-EOMCCSD and IP-EOMCCSD(3h2p) energies produced by this code against those obtained using GAMESS version 2023 R2.\cite{Gordon20_154102,Gordon23_7031} The IP-EOMCCSDT code was verified against the non-relativistic implementation in CCpy.\cite{CCpy}  All results presented in Sec.~\ref{SEC:RESULTS} were obtained from calculations that employed the frozen-core approximation, where the spinors corresponding to the [He] core of C and Cl, [Ne] core of Br, and [Ar] core of I are excluded from the CC and IP-EOMCC portions of the algorithm. Based on a set of tests using the x2c-TZVPall-2c basis set (see Table S1 in the Supplemental Information), we estimate that changes to our IP data due to the frozen-core approximation are roughly 0.005 eV, which is half the experimental uncertainty of 0.01 eV.

\section{Results and Discussion}
\label{SEC:RESULTS}

\subsection{Vertical IPs using IP-EOMCCSD and basis set effects}
\label{subsection:2h1p_basis}

We begin our discussion by analyzing the vertical IPs of the haloacetylene molecules computed at the experimentally determined geometries summarized in Table \ref{tab:hccx_geom}. The results of our 1eX2C- and DCB-X2C-IP-EOMCCSD calculations for HCCCl, HCCBr, and HCCI are shown in Figs.~\ref{fig:hcccl_ip}--\ref{fig:hcci_ip} and Tables \ref{tab:hcccl}--\ref{tab:hcci}. We also analyze the difference between IP-EOMCCSD calculations using uncontracted and recontracted basis sets; these data are summarized in Tables \ref{tab:recontraction_x2c} and \ref{tab:recontraction_ano}.

Before discussing our computed results, it is interesting to note that the experimentally determined vertical and adiabatic IP data for each of the HCCX molecules, which came from photoelectron spectroscopy measurements and are reproduced in Tables \ref{tab:hcccl}--\ref{tab:hcci}, are generally quite close to each other. Only one of the available vertical IPs differs from the corresponding adiabatic value by more than 0.1 eV (for the A state of HCCCl, see Table \ref{tab:hcccl}).  The available data for the splitting between the $\Omega = \nicefrac{3}{2}$ and $\nicefrac{1}{2}$ levels (denoted $\Delta_\Omega$) from vertical and adiabatic data are also consistent, suggesting that either set of experiments could be used as a benchmark for theory, regardless of the nature of the calculations. 
The experimental IP values also exhibit the expected progression where IP(HCCCl) > IP(HCCBr) > IP(HCCI), in accordance with the chemical intuition that it is easier to ionize electrons from shells that are further from the nucleus. The $\Omega$ splittings also show the expected pattern of $\Delta_\Omega$(HCCCl$^+$) < $\Delta_\Omega$(HCCBr$^+$) < $\Delta_\Omega$(HCCI$^+$), which is a clear indication that SOC interactions are strongest for iodine and weakest for chlorine.

Before delving into specific details for each molecule, we note some general trends. The calculated data reported in Figs.~\ref{fig:hcccl_ip}--\ref{fig:hcci_ip} and Tables \ref{tab:hcccl}--\ref{tab:hcci} show the same behavior with increasing basis set size, regardless of the identity of the molecule or the basis set family (x2c, ANO-RCC, or Dyall). Generally speaking, the vertical IP increases with the size of the basis set, and IP-EOMCCSD seems agree best with experiment when combined with double- or triple-$\zeta$-quality basis sets. The fact that large-basis IP-EOMCCSD overestimates the vertical IP suggests that IP-EOMCCSD is not converged with respect to the correlation treatment.  

\subsubsection{HCCCl}

\begin{figure*}[!htbp]
    \centering
    \includegraphics[width=\textwidth]{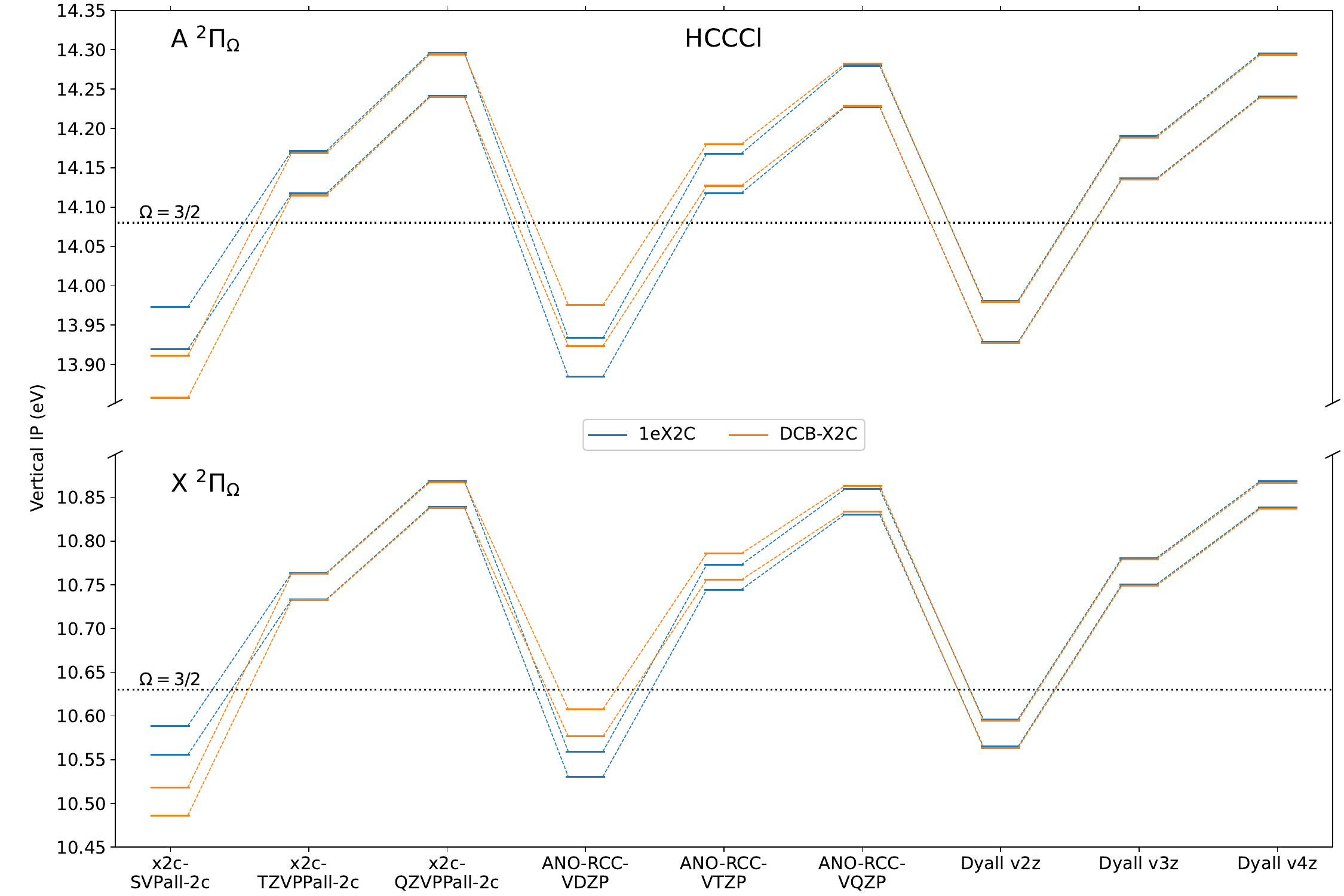}
    \caption{
        \label{fig:hcccl_ip}
        Calculated vertical IPs for HCCCl using 1eX2C- (blue) and DCB-X2C-based (orange) IP-EOMCCSD with different basis sets. Experimentally determined vertical IP values (black dotted lines) are provided for comparison.}
\end{figure*}

The computed and experimentally-obtained vertical IPs for HCCCl are provided in Fig.~\ref{fig:hcccl_ip} and Table \ref{tab:hcccl}. Note that, for this molecule, experimentally determined vertical ionization data are only available for the $\Omega=\nicefrac{3}{2}$ states, whereas the adiabatic IP data are more complete. Data from the larger molecules seem to indicate that the vertical and adiabatic $\Delta_\Omega$ splitting values are fairly consistent, though, so the adiabatic splittings for HCCCl can still serve as a benchmark for the results of the vertical IP-EOMCCSD calculations. 

\begin{table*}[!htbp]
\begin{minipage}{\linewidth}
    \caption{
        \label{tab:hcccl}
        Vertical IPs for HCCCl and spin--orbit splitting values ($\Delta_\Omega$) for HCCCl$^{+}$ states obtained from 1eX2C- and DCB-X2C-based IP-EOMCCSD calculations carried out in several basis sets. All values are reported in units of
        eV.}
    \centering
    \begin{tabular*}{\linewidth}{@{\extracolsep{\fill}} ccccccccccccc}
        \hline\hline
        \multirow{3}{*}{Basis set} & \multicolumn{6}{c}{X $^2\Pi_\Omega$} & \multicolumn{6}{c}{A $^2\Pi_\Omega$} \\
        \cline{2-7} \cline{8-13}
        & \multicolumn{2}{c}{IP ${(\Omega = \nicefrac{3}{2})}$} & \multicolumn{2}{c}{IP ${(\Omega = \nicefrac{1}{2})}$} & \multicolumn{2}{c}{$\Delta_\Omega$}
        & \multicolumn{2}{c}{IP ${(\Omega = \nicefrac{3}{2})}$} & \multicolumn{2}{c}{IP ${(\Omega = \nicefrac{1}{2})}$} & \multicolumn{2}{c}{$\Delta_\Omega$} \\
        \cline{2-3} \cline{4-5}  \cline{6-7}  \cline{8-9}  \cline{10-11} \cline{12-13}
        & 1e & DCB & 1e & DCB & 1e & DCB & 1e & DCB & 1e & DCB & 1e & DCB \\
        \hline
        x2c-SVPall-2c   & 10.56 & 10.49 & 10.59 & 10.52 & 0.03 & 0.03 & 13.92 & 13.86 & 13.97 & 13.91 & 0.05 & 0.05 \\
        x2c-TZVPPall-2c & 10.73 & 10.73 & 10.76 & 10.76 & 0.03 & 0.03 & 14.12 & 14.11 & 14.17 & 14.17 & 0.05 & 0.05 \\
        x2c-QZVPPall-2c & 10.84 & 10.84 & 10.87 & 10.87 & 0.03 & 0.03 & 14.24 & 14.24 & 14.30 & 14.29 & 0.06 & 0.05 \\
        \hline
        ANO-RCC-VDZP    & 10.53 & 10.58 & 10.56 & 10.61 & 0.03 & 0.03 & 13.88 & 13.92 & 13.93 & 13.98 & 0.05 & 0.05 \\
        ANO-RCC-VTZP    & 10.74 & 10.76 & 10.77 & 10.79 & 0.03 & 0.03 & 14.12 & 14.13 & 14.17 & 14.18 & 0.05 & 0.05 \\
        ANO-RCC-VQZP    & 10.83 & 10.83 & 10.86 & 10.86 & 0.03 & 0.03 & 14.23 & 14.23 & 14.28 & 14.28 & 0.05 & 0.05 \\
        \hline
        Dyall v2z       & 10.56 & 10.56 & 10.60 & 10.59 & 0.03 & 0.03 & 13.93 & 13.93 & 13.98 & 13.98 & 0.05 & 0.05 \\
        Dyall v3z       & 10.75 & 10.75 & 10.78 & 10.78 & 0.03 & 0.03 & 14.14 & 14.13 & 14.19 & 14.19 & 0.05 & 0.05 \\
        Dyall v4z       & 10.84 & 10.84 & 10.87 & 10.87 & 0.03 & 0.03 & 14.24 & 14.24 & 14.30 & 14.29 & 0.05 & 0.05 \\
        \hline
        Expt (vertical)\footnotemark & \multicolumn{2}{c}{10.63} & \multicolumn{2}{c}{---  } & \multicolumn{2}{c}{--- } & \multicolumn{2}{c}{14.08} & \multicolumn{2}{c}{---  } & \multicolumn{2}{c}{--- } \\
        Expt (adiabatic)\footnotemark & \multicolumn{2}{c}{10.58} & \multicolumn{2}{c}{10.60} & \multicolumn{2}{c}{0.02} & \multicolumn{2}{c}{13.87} & \multicolumn{2}{c}{13.92} & \multicolumn{2}{c}{0.05} \\
        \hline\hline
    \end{tabular*}
    \footnotetext[1]{Vertical transition from Ref.~\citenum{Kloster-Jensen70_1073}. The experimental IP values for the X and A $^2\Pi_{\nicefrac{1}{2}}$ states are not available.}
    \footnotetext[2]{Adiabatic transition from Ref.~\citenum{Maier77_1406}.}
\end{minipage}
\end{table*}

As shown in Fig.~\ref{fig:hcccl_ip} and Table \ref{tab:hcccl}, IP-EOMCCSD underestimates the experimentally determined vertical IP data by about 0.1 eV for both the X and A $^2\Pi_{\nicefrac{3}{2}}$ states, when combined with the x2c-SVPall-2c, ANO-RCC-VDZP, or Dyall v2z basis sets. The calculated vertical IPs increase by about 0.2 eV when the x2c-TZVPPall-2c, ANO-RCC-VTZP, and Dyall v3z sets are employed and by another 0.1 eV when using the x2c-QZVPPall-2c, ANO-RCC-VQZP, and Dyall v4z basis sets. For the X $^2\Pi_{\nicefrac{3}{2}}$ state of HCCCl$^+$, the double-$\zeta$ basis sets provide the best estimates of the experimentally obtained value
to within 0.7 eV or less, with 1eX2C-IP-EOMCCSD/ANO-RCC-VDZP and DCB-X2C-IP-EOMCCSD/x2c-SVPall-2c being the outliers having 0.10--0.014 eV deviations relative to experimental data. In the case of the A $^2\Pi_{\nicefrac{3}{2}}$ state, the triple-$\zeta$ basis sets provide the best estimate of the experimentally determined vertical IP value with errors in the range of 0.03--0.06 eV.
It is also interesting to note that the calculated $\Delta_\Omega$ splittings characterizing both the X and A ion states are essentially independent of the basis set and the relativistic treatment (see Table \ref{tab:hcccl}). Both 1eX2C- and DCB-X2C-IP-EOMCCSD predict the experimentally measured splitting values to within 0.01 eV, which is consistent with the experimental uncertainty range.\cite{Maier77_1406}
It is also worth noting that, as shown in Table S2 in the Supplemental Information, the IP values obtained using the cc-pVDZ-DK through cc-pVQZ-DK basis sets are very close to their analogs reported in Table \ref{tab:hcccl}. Focusing on the cc-pVQZ-DK results, the IP-EOMCCSD energetics for HCCCl are within 0.01--0.02 eV of the values evaluated using the ANO-RCC-VQZP, Dyall v4z, and x2c-QZVPPall-2c basis sets. Furthermore, the use of the cc-pV5Z-DK basis set increases the IP values by $\sim$0.04--0.05 eV compared to those computing using the cc-pVQZ-DK set. The $\Delta_\Omega$ splittings in HCCCl are also practically unaffected by the basis set size up to cc-pV5Z-DK. Thus, the IP values of HCCCl  obtained using quadruple-$\zeta$-level basis sets could be considered converged with respect to basis set size to well within 0.1 eV.

The relative behavior of 1eX2C- and DCB-X2C-IP-EOMCCSD vertical IP data in each basis set family is interesting, particularly for the smaller basis sets. The vertical IPs produced with either relativistic treatment are nearly identical when using any of the Dyall basis sets, but these values can differ substantially when using the x2c- and ANO-RCC-type sets, expecially for the smaller $\zeta$ levels. In the case of IP-EOMCCSD/x2c-SVPall-2c, 1eX2C produces vertical IP estimates that are more similar to the uncontracted Dyall v2z data, with DCB-X2C-derived IPs being lower by about 0.06--0.08 eV. This pattern is reversed when the ANO-RCC-VDZP basis set is used, with DCB-X2C producing vertical IP estimates that are closer to the uncontracted Dyall v2z values, while 1eX2C produces IPs that are lower by 0.04--0.05 eV compared to DCB-X2C. These observations reflect differences in the contraction/recontraction schemes used in each basis set family. The Dyall sets are fully uncontracted, so no recontraction scheme applies to calculations carried out with these basis sets. On the other hand, calculations using the x2c- and ANO-RCC-type use uncontracted forms of the basis sets for the 4c expansion of the Dirac Hamiltonian in 1eX2C and for the DHF procedure in DCB-X2C, after which all quantities are recontracted for use in the subsequent portions of the respective algorithms.
The difference between 1eX2C- and DCB-X2C-IP-EOMCCSD derived data becomes less apparent with increased basis set size. 1eX2C- and DCB-X2C-based results obtained in the x2c-TZVPPall-2c basis are nearly identical, but, using the ANO-RCC basis sets, good agreement between the different relativistic treatments requires the ANO-RCC-QZVP set; the ANO-RCC-VTZP basis set still produces about 0.01--0.02 eV differences. We have previously observed substantial differences in results obtained from small-basis x2c- and ANO-RCC-based calculations that include spin-orbit coupling, even in closed-shell systems.\cite{DePrinceIII24_6521} The present observations point to a similar issue in open-shell molecules where SOC effects are more relevant.
We will return to this recontraction scheme issue later.

\subsubsection{HCCBr}

\begin{figure*}[!htbp]
    \centering
    \includegraphics[width=\textwidth]{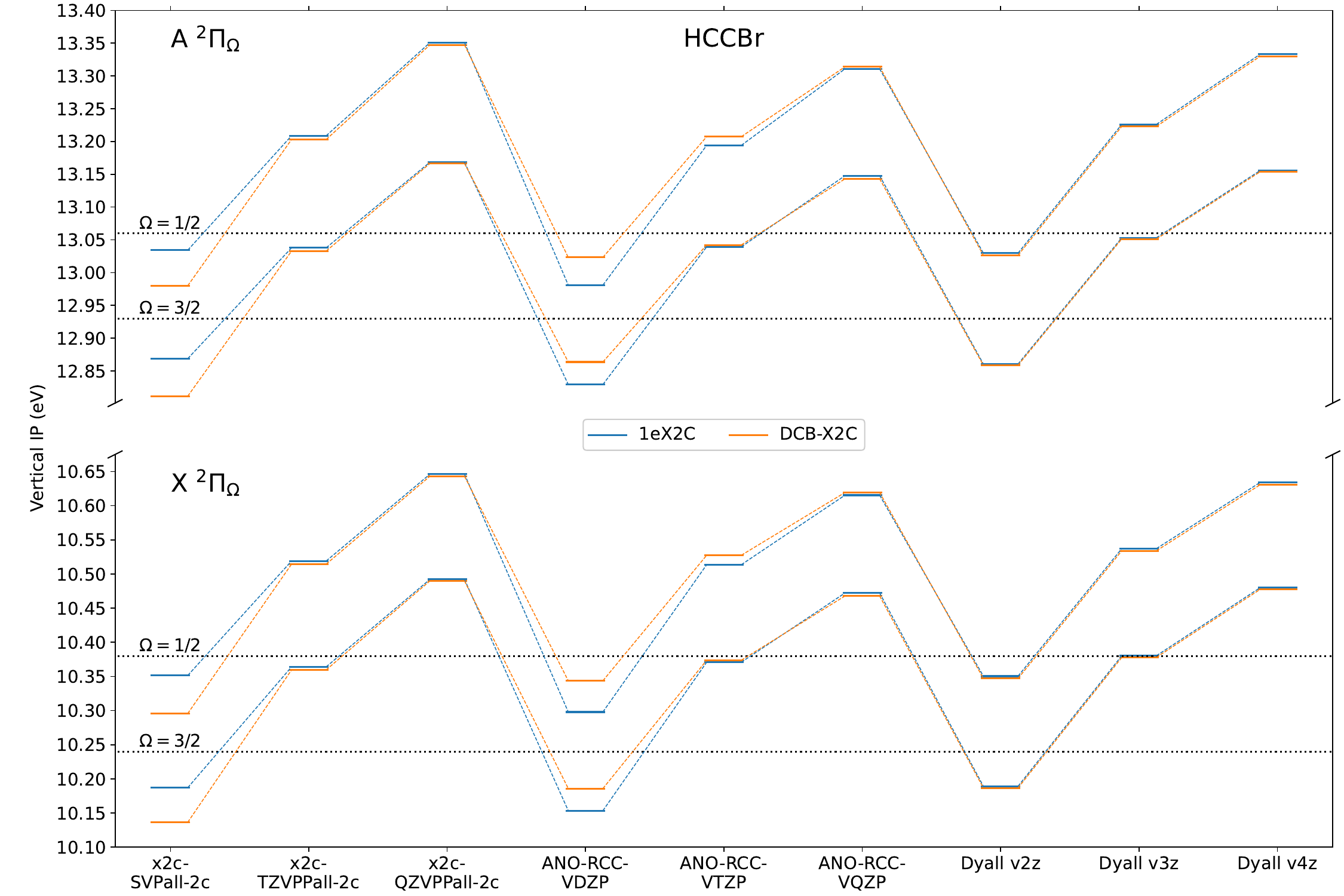}
    \caption{
        \label{fig:hccbr_ip}
        Calculated vertical IPs for HCCBr using 1eX2C- (blue) and DCB-X2C-based (orange) IP-EOMCCSD with different basis sets. Experimentally determined vertical IP values (black dotted lines) are provided for comparison.}
\end{figure*}

Computed and experimentally-obtained vertical IPs for HCCBr are provided in Fig.~\ref{fig:hccbr_ip} and Table \ref{tab:hccbr}. As can be seen in Table \ref{tab:hccbr}, more complete vertical IP data are provided in this case, as compared to HCCCl. The data in Table \ref{tab:hccbr} also demonstrate that experimentally-obtained vertical and adiabatic IP values are in reasonable agreement, to within 0.06--0.07 eV. However, we suspect that the adiabatic IP values are more reliable than the vertical IP values, for the following reasons. First, the experimental data for the X $^2\Pi_\Omega$ states seem to be inconsistent because the adiabatic IP values are larger than the vertical ones, whereas geometric relaxation effects should lead to the opposite trend. Second, Ref.~\citenum{Maier77_1406} notes challenges in assigning the vertical IPs for the A $^2\Pi_{\nicefrac{3}{2}}$ and $^2\Pi_{\nicefrac{1}{2}}$ states because the splitting is similar in magnitude to the vibrational spacings for the molecules. 

\begin{table*}[!htbp]
\begin{minipage}{\linewidth}
    \caption{
        \label{tab:hccbr}
        Vertical IPs for HCCBr and spin--orbit splitting values ($\Delta_\Omega$) for HCCBr$^{+}$ states obtained from 1eX2C- and DCB-X2C-based IP-EOMCCSD calculations carried out in several basis sets. All values are reported in units of
        eV.}
    \centering
    \begin{tabular*}{\linewidth}{@{\extracolsep{\fill}} ccccccccccccc}
        \hline\hline
        \multirow{3}{*}{Basis set} & \multicolumn{6}{c}{X $^2\Pi_\Omega$} & \multicolumn{6}{c}{A $^2\Pi_\Omega$} \\
        \cline{2-7} \cline{8-13}
        & \multicolumn{2}{c}{IP ${(\Omega = \nicefrac{3}{2})}$} & \multicolumn{2}{c}{IP ${(\Omega = \nicefrac{1}{2})}$} & \multicolumn{2}{c}{$\Delta_\Omega$}
        & \multicolumn{2}{c}{IP ${(\Omega = \nicefrac{3}{2})}$} & \multicolumn{2}{c}{IP ${(\Omega = \nicefrac{1}{2})}$} & \multicolumn{2}{c}{$\Delta_\Omega$} \\
        \cline{2-3} \cline{4-5}  \cline{6-7}  \cline{8-9}  \cline{10-11} \cline{12-13}
        & 1e & DCB & 1e & DCB & 1e & DCB & 1e & DCB & 1e & DCB & 1e & DCB \\
        \hline
        x2c-SVPall-2c   & 10.19 & 10.14 & 10.35 & 10.30 & 0.16 & 0.16 & 12.87 & 12.81 & 13.03 & 12.98 & 0.17 & 0.17 \\
        x2c-TZVPPall-2c & 10.36 & 10.36 & 10.52 & 10.51 & 0.16 & 0.15 & 13.04 & 13.03 & 13.21 & 13.20 & 0.17 & 0.17 \\
        x2c-QZVPPall-2c & 10.49 & 10.49 & 10.65 & 10.64 & 0.15 & 0.15 & 13.17 & 13.17 & 13.35 & 13.35 & 0.18 & 0.18 \\
        \hline
        ANO-RCC-VDZP    & 10.15 & 10.19 & 10.30 & 10.34 & 0.14 & 0.16 & 12.83 & 12.86 & 12.98 & 13.02 & 0.15 & 0.16 \\
        ANO-RCC-VTZP    & 10.37 & 10.37 & 10.51 & 10.53 & 0.14 & 0.15 & 13.04 & 13.04 & 13.19 & 13.21 & 0.15 & 0.17 \\
        ANO-RCC-VQZP    & 10.47 & 10.47 & 10.62 & 10.62 & 0.14 & 0.15 & 13.15 & 13.14 & 13.31 & 13.31 & 0.16 & 0.17 \\
        \hline
        Dyall v2z       & 10.19 & 10.19 & 10.35 & 10.35 & 0.16 & 0.16 & 12.86 & 12.86 & 13.03 & 13.03 & 0.17 & 0.17 \\
        Dyall v3z       & 10.38 & 10.38 & 10.54 & 10.53 & 0.16 & 0.16 & 13.05 & 13.05 & 13.23 & 13.22 & 0.17 & 0.17 \\
        Dyall v4z       & 10.48 & 10.48 & 10.63 & 10.63 & 0.15 & 0.15 & 13.16 & 13.15 & 13.33 & 13.33 & 0.18 & 0.18 \\
        \hline
        Expt (vertical)\footnotemark & \multicolumn{2}{c}{10.24} & \multicolumn{2}{c}{10.38} & \multicolumn{2}{c}{0.14} & \multicolumn{2}{c}{12.93} & \multicolumn{2}{c}{13.06} & \multicolumn{2}{c}{0.13} \\
        Expt (adiabatic)\footnotemark & \multicolumn{2}{c}{10.31} & \multicolumn{2}{c}{10.44} & \multicolumn{2}{c}{0.13} & \multicolumn{2}{c}{12.86} & \multicolumn{2}{c}{13.06} & \multicolumn{2}{c}{0.20} \\
        \hline\hline
    \end{tabular*}
    \footnotetext[1]{Vertical transition from Ref.~\citenum{Kloster-Jensen70_1073}.}
    \footnotetext[2]{Adiabatic transition from Ref.~\citenum{Maier77_1406}.}
\end{minipage}
\end{table*}

As for the computed data, we find once again that the vertical IP values are underestimated when using small basis sets, and the predicted IP values increase as the basis set improves from double- to triple-$\zeta$ quality (0.20--0.25 eV for the X $^2\Pi_\Omega$ states) and from triple- to quadruple-$\zeta$ quality (0.10--0.15 eV, again, for the X $^2\Pi_\Omega$ states). These increases are slightly larger than those we observed for HCCCl. 
The calculated vertical IP values for the A $^2\Pi_\Omega$ states produce slightly smaller errors (\emph{cf.} Table \ref{tab:hccbr}). Similar to the HCCCl molecule, the double-$\zeta$ basis sets produce the best vertical IP estimates relative to experimentally determined values for both X and A $^2\Pi_\Omega$ states.
As was the case for HCCCl, the good performance of small basis sets is simply due to a cancellation of errors, and the large errors associated with calculations carried out in quadruple-$\zeta$ basis sets is mainly due to correlation effects that are missing at the IP-EOMCCSD level of theory. It is also noteworthy that the  cc-pV$n$Z-DK ($n={}$D, T, Q) data provided in the Supplementary Information for HCCBr are, again, very similar to their counterparts in Table \ref{tab:hccbr}, with the IP values and $\Delta_\Omega$ splittings computed using the cc-pVQZ-DK basis set being within 0.01--0.03 eV of the values computed using the ANO-RCC-VQZP and Dyall v4z basis sets. The discrepancies between IP values obtained using the cc-pVQZ-DK and x2c-QZVPPall-2c basis sets are slightly larger, between 0.03--0.05 eV, but both basis sets still produce $\Delta_\Omega$ splittings that are within 0.01--0.02 eV of each other. Similar to the smaller HCCCl case, going from the cc-pVQZ-DK basis to the cc-pV5Z-DK basis increases the IP values of HCCBr by about 0.05 eV (see Table S2 in the Supplemental Information).

Now, consider the spin-orbit splittings, $\Delta_\Omega$. For HCCBr, IP-EOMCCSD predicts that $\Delta_\Omega$ for the X $^2\Pi_\Omega$ states are lower than $\Delta_\Omega$ for the A $^2\Pi_\Omega$ states, which is consistent with $\Delta_\Omega$ values derived from adiabatic IP data but not with the values from vertical IP data. The errors in the computed $\Delta_\Omega$ values (0.01--0.03 eV for the X states and 0.02--0.05 eV for the A states) are slighly larger than the errors $\Delta_\Omega$ observed in the case of HCCCl. Recovering the 0.01 eV errors observed for HCCCl will require the consideration of higher-order correlation effects, as would be captured by IP-EOMCCSD(3h2p) or IP-EOMCCSDT (see the analysis for adiabatic IPs below). Nevertheless, it is encouraging to see that the 1eX2C- and DCB-X2C-IP-EOMCCSD splittings are within 0.01 eV from each other regardless of the basis set, which suggests that the DCB-parameterized SNSO factor employed in the 1eX2C calculation yields SOC splittings that are competitive in accuracy with those from the more complete DCB-X2C framework. We also note $\Delta_\Omega$ values are not too sensitive to the quality of the basis set, varying by no more than 0.01 eV as we increase the $\zeta$-level within a given basis set family, for a specific relativistic treatment.

As was the case with HCCCl, IP values obtained using uncontracted versus contracted basis sets for HCCBr can differ significantly, depending on the relativistic treatment. Again, 1eX2C- and DCB-X2C-IP-EOMCCSD calculations that use the uncontracted Dyall basis sets produce practically identical IP values. 1eX2C-IP-EOMCCSD/x2c-SVPall-2c predicts vertical IP values that are within 0.01 eV of the uncontracted Dyall v2z results, for both of the ionized states we consider. In contrast, DCB-X2C-IP-EOMCCSD/x2c-SVPall-2c yields IP values that are lower by about 0.05 eV compared to their 1eX2C counterparts. As was observed for HCCCl, this behavior flips when we consider the results of calculations carried out using the ANO-RCC-VDZP basis set. In this case, DCB-X2C-IP-EOMCCSD generates vertical IPs that are more similar to those obtained using the uncontracted Dyall v2z basis set, while vertical IPs from 1eX2C-IP-EOMCCSD calculations are lower.

\subsubsection{HCCI}

\begin{figure*}[!htbp]
    \centering
    \includegraphics[width=\textwidth]{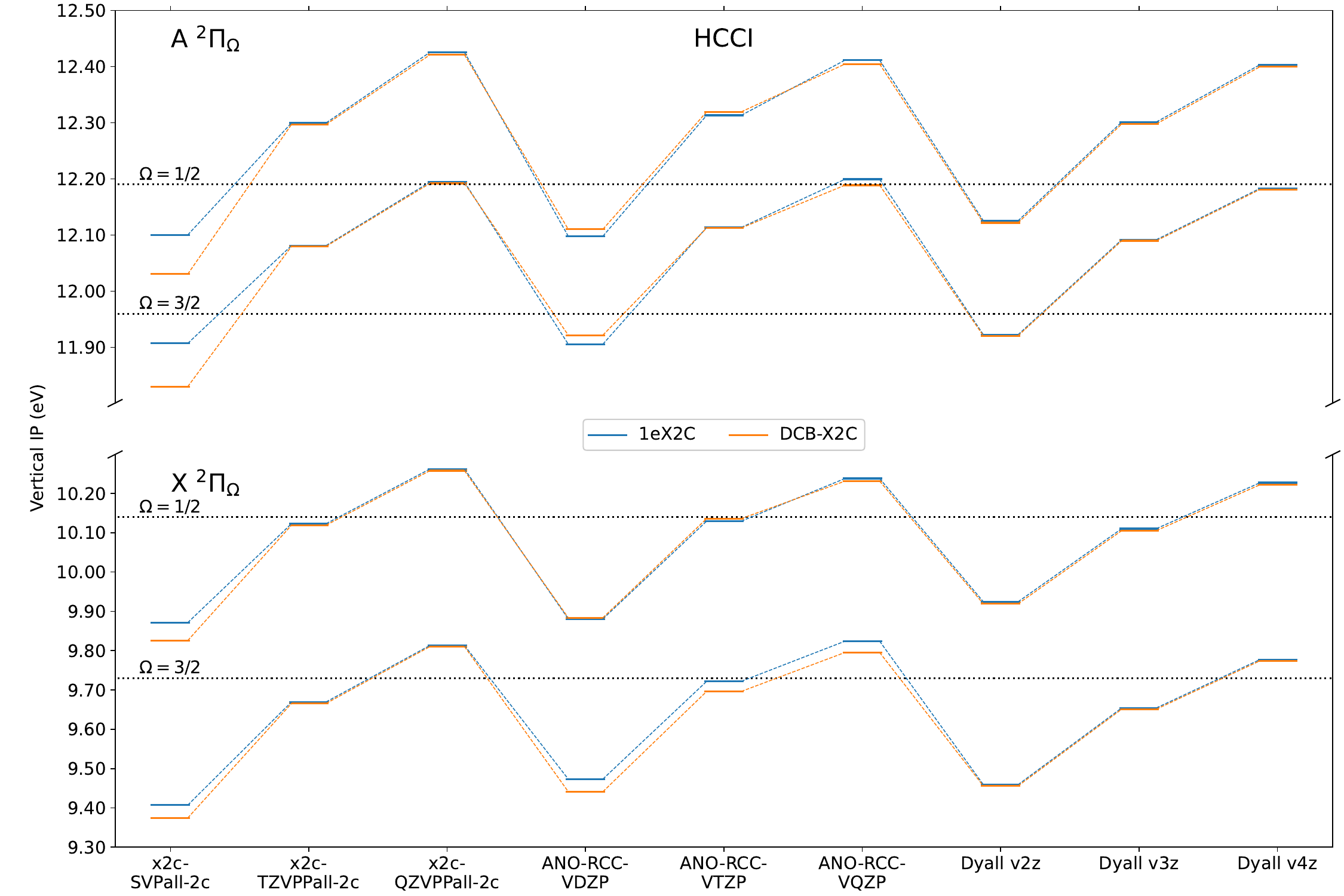}
    \caption{
        \label{fig:hcci_ip}
        Calculated vertical IPs for HCCI using 1eX2C- (blue) and DCB-X2C-based (orange) IP-EOMCCSD with different basis sets. Experimentally determined vertical IP values (black dotted lines) are provided for comparison.}
\end{figure*}

Computed and experimentally-obtained vertical IPs for HCCI are provided in Fig.~\ref{fig:hcci_ip} and Table \ref{tab:hcci}. Of the three cations studied in this work, HCCI$^+$ is characterized by the largest SOC splittings in the X $^2\Pi_\Omega$ ($\sim$0.4 eV) and A $^2\Pi_\Omega$ ($\sim$0.25 eV) states. As such, this system serves as a good benchmark for approaches that account for spin-dependent relativistic effects. Like HCCBr, the experimental data for the vertical and adiabatic IPs of HCCI are more complete than for HCCCl. However, unlike HCCBr, the relative magnitudes of the vertical and adiabatic IPs make physical sense for HCCI, with the adiabatic IPs having the lower values, for both the X and A $^2\Pi_\Omega$ states.
As a result, we should be able to directly compare the quality of our calculated vertical IP data with the experimentally determined values from Ref.~\citenum{Kloster-Jensen70_1073}.

The DCB-X2C-IP-EOMCCSD/x2c-QZVPPall-2c calculations on HCCI revealed a linear dependency issue, which we attribute to one of the exponents in the contraction of 6 p-type primitives (12.459849670) being  close to the exponent of another uncontracted primitive (12.560234085). To resolve this issue, 
we removed the second primitive. In the Chronus Quantum package, this is achieved by changing the second exponent to 12.459849670. In the DHF portion of this program, primitives belonging to the same angular momentum shell that have the same exponent are interpreted as redundant, and only one is used in the calculation. Upon recontraction of the basis after the SCF step, though, both primitive functions are reintroduced. In this way, the correlated portion of the calculation retains the same dimensionality as in the 1eX2C/x2c-QZVPPall-2c calculations on HCCI, for which there were no linear dependency issues.

\begin{table*}[!htbp]
\begin{minipage}{\linewidth}
    \caption{
        \label{tab:hcci}
        Vertical IPs for HCCI and spin--orbit splitting values ($\Delta_\Omega$) for HCCI$^{+}$ states obtained from 1eX2C- and DCB-X2C-based IP-EOMCCSD calculations carried out in several basis sets. All values are reported in units of eV.}
    \centering
    \begin{tabular*}{\linewidth}{@{\extracolsep{\fill}} ccccccccccccc}
        \hline\hline
        \multirow{3}{*}{Basis set} & \multicolumn{6}{c}{X $^2\Pi_\Omega$} & \multicolumn{6}{c}{A $^2\Pi_\Omega$} \\
        \cline{2-7} \cline{8-13}
        & \multicolumn{2}{c}{IP ${(\Omega = \nicefrac{3}{2})}$} & \multicolumn{2}{c}{IP$ {(\Omega = \nicefrac{1}{2})}$} & \multicolumn{2}{c}{$\Delta_\Omega$}
        & \multicolumn{2}{c}{IP ${(\Omega = \nicefrac{3}{2})}$} & \multicolumn{2}{c}{IP$ {(\Omega = \nicefrac{1}{2})}$} & \multicolumn{2}{c}{$\Delta_\Omega$} \\
        \cline{2-3} \cline{4-5}  \cline{6-7}  \cline{8-9}  \cline{10-11} \cline{12-13}
        & 1e & DCB & 1e & DCB & 1e & DCB & 1e & DCB & 1e & DCB & 1e & DCB \\
        \hline
        x2c-SVPall-2c   &  9.41 &  9.37 &  9.87 &  9.83 & 0.46 & 0.45 & 11.91 & 11.83 & 12.10 & 12.03 & 0.19 & 0.20 \\
        x2c-TZVPPall-2c &  9.67 &  9.67 & 10.12 & 10.12 & 0.45 & 0.45 & 12.08 & 12.08 & 12.30 & 12.30 & 0.22 & 0.22 \\
        x2c-QZVPPall-2c &  9.81 &  9.81 & 10.26 & 10.26 & 0.45 & 0.45 & 12.19 & 12.19 & 12.43 & 12.42 & 0.23 & 0.23 \\
        \hline
        ANO-RCC-VDZP    &  9.47 &  9.44 &  9.88 &  9.88 & 0.41 & 0.44 & 11.91 & 11.92 & 12.10 & 12.11 & 0.19 & 0.19 \\
        ANO-RCC-VTZP    &  9.72 &  9.70 & 10.13 & 10.14 & 0.41 & 0.44 & 12.11 & 12.11 & 12.31 & 12.32 & 0.20 & 0.21 \\
        ANO-RCC-VQZP    &  9.82 &  9.80 & 10.24 & 10.23 & 0.41 & 0.44 & 12.20 & 12.19 & 12.41 & 12.40 & 0.21 & 0.22 \\
        \hline
        Dyall v2z       &  9.46 &  9.46 &  9.92 &  9.92 & 0.47 & 0.46 & 11.92 & 11.92 & 12.13 & 12.12 & 0.20 & 0.20 \\
        Dyall v3z       &  9.65 &  9.65 & 10.11 & 10.11 & 0.46 & 0.45 & 12.09 & 12.09 & 12.30 & 12.30 & 0.21 & 0.21 \\
        Dyall v4z       &  9.78 &  9.77 & 10.23 & 10.22 & 0.45 & 0.45 & 12.18 & 12.18 & 12.40 & 12.40 & 0.22 & 0.22 \\
        \hline
        Expt (vertical)\footnotemark & \multicolumn{2}{c}{9.73} & \multicolumn{2}{c}{10.14} & \multicolumn{2}{c}{0.41} & \multicolumn{2}{c}{11.96} & \multicolumn{2}{c}{12.19} & \multicolumn{2}{c}{0.23} \\
        Expt (adiabatic)\footnotemark & \multicolumn{2}{c}{9.71} & \multicolumn{2}{c}{10.11} & \multicolumn{2}{c}{0.40} & \multicolumn{2}{c}{11.87} & \multicolumn{2}{c}{12.12} & \multicolumn{2}{c}{0.25} \\
        \hline\hline
    \end{tabular*}
    \footnotetext[1]{Vertical transition from Ref.~\citenum{Kloster-Jensen70_1073}.}
    \footnotetext[2]{Adiabatic transition from Ref.~\citenum{Maier77_1406}.}
\end{minipage}
\end{table*}

The basis-set dependence of the calculated vertical IPs for HCCI resembles that observed for HCCCl and HCCBr. The IP values increase with the $\zeta$-level of the basis sets, by $\sim$0.2--0.3 eV when going from double- to triple-$\zeta$, followed by another $\sim$0.1 eV when going from triple- to quadruple-$\zeta$. This convergence behavior is similar across basis set families. We also see a pattern that resembles the one observed in the other systems, namely, that the vertical IPs are best reproduced in smaller basis sets. In this case, the description of the X $^2\Pi_\Omega$ states is best using the triple-$\zeta$ basis sets while the A $^2\Pi_\Omega$ states can be reproduced accurately using double-$\zeta$-quality sets. Again, this agreement is due to a cancellation of error, and the remaining error from calculations carried out in the largest basis sets can mainly be attributed to missing correlation effects.

Computed splitting values ($\Delta_\Omega$) vary by 0.01 to 0.04 eV, depending on the size and type of basis set employed. Interestingly, $\Delta_\Omega$ values for the the X $^2\Pi_\Omega$ states are fairly consistent across $\zeta$ levels, within the x2c and ANO-RCC basis set families, varying by at most 0.01 eV, whereas slightly larger variations are observed for values computed using the Dyall sets (up to 0.02 eV). On the other hand, $\Delta_\Omega$ values for the the A $^2\Pi_\Omega$ states display a somewhat larger dependence on the $\zeta$ level, increasing by 0.02--0.04 eV going from double- to triple- to quadruple-$\zeta$ quality, within a given basis set family. As for the different relativistic treatments, $\Delta_\Omega$ values computed using 1eX2C- and DCB-X2C-IP-EOMCCSD agree to within 0.01 eV, with the exception of the 0.03 eV differences we observe for the X $^2\Pi_\Omega$ states, as described by the ANO-RCC basis sets.

As was observed for the smaller molecules, IP values for HCCI derived from calculations that use uncontracted vs contracted basis sets can differ significantly, depending on the relativistic treatment. Again, IP estimates obtained using the Dyall basis sets are insensitive to the relativistic treatment, while those derived from calculations using the x2c- and ANO-RCC-type basis sets may differ, and these differences are largest in the double-$\zeta$-quality basis sets. Within the x2c-SVPall-2c basis set, 1eX2C-derived values agree best with Dyall v2z results, whereas DCB-X2C-based IPs are lower than the 1eX2C-based ones by about 0.04 eV and 0.07--0.08 eV in the X and A $^2\Pi_\Omega$ states, respectively. The IPs computed using the ANO-RCC-VDZP basis set behave slightly differently than was observed for the smaller molecules. The vertical IP estimate for the X $^2\Pi_{\nicefrac{3}{2}}$ state computed using DCB-X2C-IP-EOMCCSD is lower than the 1eX2C analog, and the X $^2\Pi_{\nicefrac{1}{2}}$ IP values resulting from 1eX2C- and DCB-X2C-IP-EOMCCSD agree to within less than 0.01 eV. In the other molecules, 1eX2C consistently produced the lower IP values when using this basis. On the other hand, for the A $^2\Pi_\Omega$ states, the relative ordering of the IP values from 1eX2C- and DCB-X2C-IP-EOMCCSD agrees with those observed for the other molecules. Once we employ larger basis sets, 1eX2C- and DCB-X2C-IP-EOMCCSD, for the most part, produce more similar vertical IP values, although some small differences persist even when using the ANO-RCC-QZVP set.

\subsubsection{Basis set recontraction with 1eX2C and DCB-X2C}

As discussed above, vertical IPs computed in small contracted basis sets show a surprising and non-negligible sensitivity to the choice of relativistic treatment. We have performed additional calculations using the fully uncontracted x2c-SVPall-2c and ANO-RCC-VDZP basis sets, the results of which are reported in Tables \ref{tab:recontraction_x2c} and \ref{tab:recontraction_ano}, respectively, in order to understand this behavior. In short, differences between results obtained using 1eX2C- and DCB-X2C-IP-EOMCC in small contracted basis sets can be attributed to a recontraction error introduced at the 4c to 2c transformation.

\begin{table*}[!htbp]
\begin{minipage}{\linewidth}
    \caption{
        \label{tab:recontraction_x2c}
        Comparison of vertical IPs obtained using IP-EOMCCSD with recontracted and uncontracted x2c-SVPall-2c basis sets, in conjunction with the 1eX2C (1e) and DCB-X2C (DCB) mean-field treatments. All values are reported in eV.}
    \centering
    \begin{tabular*}{\linewidth}{@{\extracolsep{\fill}} ccrrrrrr}
        \hline\hline
        \multirow{2}{*}{Cation} & \multirow{2}{*}{State} & \multicolumn{3}{c}{IP (recontracted)\footnotemark[1]} & \multicolumn{3}{c}{IP (uncontracted)} \\
        \cline{3-5} \cline{6-8}
        & & 1e & DCB & $\mathrm{1e}-\mathrm{DCB}$ & 1e & DCB & $\mathrm{1e}-\mathrm{DCB}$ \\
        \hline
        \multirow{4}{*}{HCCCl$^+$}
        & X $^2\Pi_{\nicefrac{3}{2}}$ & 10.556 & 10.486 & 0.069 & 10.519 & 10.517 & 0.002 \\
        & X $^2\Pi_{\nicefrac{1}{2}}$ & 10.589 & 10.518 & 0.071 & 10.550 & 10.548 & 0.002 \\
        & A $^2\Pi_{\nicefrac{3}{2}}$ & 13.919 & 13.858 & 0.062 & 13.882 & 13.881 & 0.001 \\
        & A $^2\Pi_{\nicefrac{1}{2}}$ & 13.973 & 13.911 & 0.062 & 13.935 & 13.933 & 0.002 \\
        \hline
        \multirow{4}{*}{HCCBr$^+$}
        & X $^2\Pi_{\nicefrac{3}{2}}$ & 10.187 & 10.136 & 0.051 & 10.174 & 10.171 & 0.003 \\
        & X $^2\Pi_{\nicefrac{1}{2}}$ & 10.351 & 10.296 & 0.056 & 10.334 & 10.330 & 0.004 \\
        & A $^2\Pi_{\nicefrac{3}{2}}$ & 12.869 & 12.812 & 0.057 & 12.842 & 12.840 & 0.002 \\
        & A $^2\Pi_{\nicefrac{1}{2}}$ & 13.034 & 12.980 & 0.054 & 13.012 & 13.009 & 0.003 \\
        \hline
        \multirow{4}{*}{HCCI$^+$}
        & X $^2\Pi_{\nicefrac{3}{2}}$ &  9.407 &  9.374 & 0.033 &  9.415 &  9.412 & 0.003 \\
        & X $^2\Pi_{\nicefrac{1}{2}}$ &  9.871 &  9.826 & 0.046 &  9.878 &  9.873 & 0.005 \\
        & A $^2\Pi_{\nicefrac{3}{2}}$ & 11.907 & 11.830 & 0.077 & 11.864 & 11.862 & 0.002 \\
        & A $^2\Pi_{\nicefrac{1}{2}}$ & 12.100 & 12.031 & 0.070 & 12.072 & 12.069 & 0.004 \\
        \hline\hline
    \end{tabular*}
    \footnotetext[1]{See Tables \ref{tab:hcccl}--\ref{tab:hcci}.}
\end{minipage}
\end{table*}

\begin{table*}[!htbp]
\begin{minipage}{\linewidth}
    \caption{
        \label{tab:recontraction_ano}
        Comparison of vertical IPs obtained using IP-EOMCCSD with recontracted and uncontracted ANO-RCC-VDZP basis sets, in conjunction with the 1eX2C (1e) and DCB-X2C (DCB) mean-field treatments. All values are reported in eV.}
    \centering
    \begin{tabular*}{\linewidth}{@{\extracolsep{\fill}} ccrrrrrr}
        \hline\hline
        \multirow{2}{*}{Cation} & \multirow{2}{*}{State} & \multicolumn{3}{c}{IP (recontracted)\footnotemark[1]} & \multicolumn{3}{c}{IP (uncontracted)} \\
        \cline{3-5} \cline{6-8}
        & & 1e & DCB & $\mathrm{1e}-\mathrm{DCB}$ & 1e & DCB & $\mathrm{1e}-\mathrm{DCB}$ \\
        \hline
        \multirow{4}{*}{HCCCl$^+$}
        & X $^2\Pi_{\nicefrac{3}{2}}$ & 10.531 & 10.577 & $-0.046$ & 10.695 & 10.693 & 0.001 \\
        & X $^2\Pi_{\nicefrac{1}{2}}$ & 10.559 & 10.608 & $-0.048$ & 10.726 & 10.724 & 0.002 \\
        & A $^2\Pi_{\nicefrac{3}{2}}$ & 13.885 & 13.924 & $-0.039$ & 14.061 & 14.060 & 0.001 \\
        & A $^2\Pi_{\nicefrac{1}{2}}$ & 13.934 & 13.975 & $-0.042$ & 14.115 & 14.113 & 0.002 \\
        \hline
        \multirow{4}{*}{HCCBr$^+$}
        & X $^2\Pi_{\nicefrac{3}{2}}$ & 10.153 & 10.186 & $-0.032$ & 10.300 & 10.298 & 0.002 \\
        & X $^2\Pi_{\nicefrac{1}{2}}$ & 10.298 & 10.344 & $-0.046$ & 10.460 & 10.457 & 0.003 \\
        & A $^2\Pi_{\nicefrac{3}{2}}$ & 12.829 & 12.864 & $-0.034$ & 12.971 & 12.969 & 0.002 \\
        & A $^2\Pi_{\nicefrac{1}{2}}$ & 12.981 & 13.023 & $-0.042$ & 13.141 & 13.138 & 0.003 \\
        \hline
        \multirow{4}{*}{HCCI$^+$}
        & X $^2\Pi_{\nicefrac{3}{2}}$ &  9.473 &  9.441 &   0.032  &  9.716 &  9.713 & 0.003 \\
        & X $^2\Pi_{\nicefrac{1}{2}}$ &  9.881 &  9.884 & $-0.003$ & 10.162 & 10.157 & 0.005 \\
        & A $^2\Pi_{\nicefrac{3}{2}}$ & 11.906 & 11.922 & $-0.016$ & 12.090 & 12.088 & 0.002 \\
        & A $^2\Pi_{\nicefrac{1}{2}}$ & 12.098 & 12.110 & $-0.012$ & 12.322 & 12.319 & 0.004 \\
        \hline\hline
    \end{tabular*}
    \footnotetext[1]{See Tables \ref{tab:hcccl}--\ref{tab:hcci}.}
\end{minipage}
\end{table*}

IP-EOMCCSD data obtained using recontracted and uncontracted x2c-SVPall-2c basis sets are tabulated in Table \ref{tab:recontraction_x2c}. IP values obtained with 1eX2C- and DCB-X2C-IP-EOMCCSD and the  contracted x2c-SVPall-2c set differ by about 0.06--0.07 eV for HCCCl, 0.05--0.06 eV for HCCBr, and $\sim$0.04 eV for the X state and $\sim$0.07 eV for the A state in the case of HCCI (cf.~Tables \ref{tab:hcccl}--\ref{tab:hcci}). These discrepancies decrease by an order of magnitude when the uncontracted x2c-SVPall-2c basis set is employed. For example, the largest difference of 0.005 eV, for the X $^2\Pi_{\nicefrac{1}{2}}$ state of HCCI, is roughly 9 times smaller than the 0.046 eV difference observed for the IP values obtained with recontracted basis set.

Analogous data generated using the ANO-RCC-VDZP basis set are presented in Table \ref{tab:recontraction_ano}, where we again see improved agreement between IPs from 1eX2C- and DCB-X2C-IP-EOMCCSD when the full calculations are carried out in the uncontracted basis sets. Indeed, while results obtained from the two relativistic treatments and the recontracted ANO-RCC-VDZP basis set differ by about 0.03--0.05 eV, for the HCCCl$^+$ and HCCBr$^+$ cations, this difference reduces to only 0.001--0.002 eV if the calculations are carried out in the uncontracted form of the basis. 
Moreover, for HCCI, in particular, calculations carried out using the recontracted basis produce differences between 1eX2C- and DCB-X2C-based results that vary dramatically, depending on the state. The smallest discrepancy is $-0.003$ eV for the X $^2\Pi_{\nicefrac{1}{2}}$ state, the largest one is 0.032 eV for the X $^2\Pi_{\nicefrac{3}{2}}$, and the A $^2\Pi_\Omega$ states show differences in the 0.01--0.02 eV range. The use of the uncontracted basis throughout the full calculation reduces these differences to 0.002--0.005 eV range, for all states. These data confirm that the principal source of the quantitative disagreement between small-basis 1eX2C- and DCB-X2C-IP-EOMCCSD is the recontraction of the basis after the 4c to 2c transformation.

\subsection{Correlation effects beyond IP-EOMCCSD}
\label{subsection:correlation}

The preceding analyses demonstrate that X2C-IP-EOMCCSD calculations carried out in large (quadruple-$\zeta$-quality) basis sets do not provide accurate predictions of experimentally-obtained IP values for the molecules considered in this work. IP values are overestimated by about 0.2--0.3 eV in these calculations. The poor agreement with experiment does not stem from choosing the 1eX2C- versus DCB-X2C-HF reference, as computed IP values from both of these frameworks agree quite well in the large-basis-set limit. Hence, remaining errors are due to a lack of higher-order correlation effects that are missing in IP-EOMCCSD. In this section, we consider two approaches that go beyond the correlation level captured by IP-EOMCCSD: the IP-EOMCCSD(3h2p) approach, which augments the EOM operator ($\hat{R}_K$) but does not change the correlation treatment of the ground state, and IP-EOMCCSDT, which increases the correlation level in both $\hat{R}_K$ and $\hat{T}$. 

In the following analysis, we apply both IP-EOMCCSD and higher-order correlation treatments to the adiabatic, rather than vertical, IPs for the HCCX molecules. We focus on the adiabatic IPs because the available experimental data in Ref.~\citenum{Maier77_1406} are more recent and complete than those for the vertical IPs reported in Ref.~\citenum{Kloster-Jensen70_1073}. We also limit our study to DCB-X2C-based treatments of relativistic effects in order to minimize possible sources of error. In the interest of computational efficiency, we wish to evaluate the $\mathscr{O} (\mathscr{N}^8)$-scaling CCSDT and $\mathscr{O} (\mathscr{N}^7)$-scaling IP-EOMCCSDT/IP-EOMCCSD(3h2p) portions of the algorithm using a contracted basis set.  From our vertical IP analysis, we recall that, of the two contracted basis set families, the ANO-RCC family provides the best agreement with results obtained using Dyall-type basis sets when the relativistic treatment is DCB-X2C. As such, we do not consider the x2c-type basis sets below.

We adopt the following composite protocols to account for higher-order correlation effects missing at the IP-EOMCCSD level of theory. First, to obtain IP values for state $K$ that approximate those from IP-EOMCCSDT, we define
\begin{equation}
\label{eqn:composite_ip1}
    \mathrm{IP}_K = \mathrm{IP}_K^\mathrm{SD/QZ}
                    + \left[\mathrm{IP}_K^\mathrm{SDT/DZ}
                    - \mathrm{IP}_K^\mathrm{SD/DZ}\right],
\end{equation}
where the first term on the right-hand side of the equation is the adiabatic IP value obtained using DCB-X2C-IP-EOMCCSD/ANO-RCC-VQZP, and the next two terms correspond to the difference between DCB-X2C-IP-EOMCCSDT and DCB-X2C-IP-EOMCCSD adiabatic IP values computed using the smaller ANO-RCC-VDZP basis set.  
Second, to obtain IP values that approximate those from IP-EOMCCSD(3h2p), we define
\begin{equation}
\label{eqn:composite_ip2}
    \mathrm{IP}_K^\prime = \mathrm{IP}_K^\mathrm{SD/QZ}
                    + \left[\mathrm{IP}_K^\mathrm{SD(3h2p)/DZ}
                    - \mathrm{IP}_K^\mathrm{SD/DZ}\right],
\end{equation}
where $\mathrm{IP}_K^\mathrm{SD(3h2p)/DZ}$ refers to the adabatic IP value obtained from an IP-EOMCCSD(3h2p) calculation carried out in the ANO-RCC-VDZP basis set.

The equilibrium geometries of the neutral and two lowest ion states, as summarized in Table \ref{tab:hccx_geom_adiabatic}, all have $C_{\infty v}$ symmetry.\cite{Schlegel24_117313} The vibrational analysis in Ref.~\citenum{Schlegel24_117313} indicates that the difference between the zero-point vibrational energies of the neutral molecule and cation in either the X $^2\Pi_\Omega$ and A $^2\Pi_\Omega$ states are on the order of only 100 cm$^{-1}$ or around 0.01 eV. Given that the experimental uncertainty for the adiabatic 0--0 transition reported in Ref.~\citenum{Maier77_1406} is also 0.01 eV, the computed electronic A--X adiabatic transition energy, $\omega_\mathrm{A-X}$, should be an acceptable approximation to the experimentally-obtained value. 

\begin{figure}[!htbp]
    \centering
    \includegraphics[width=0.8\columnwidth]{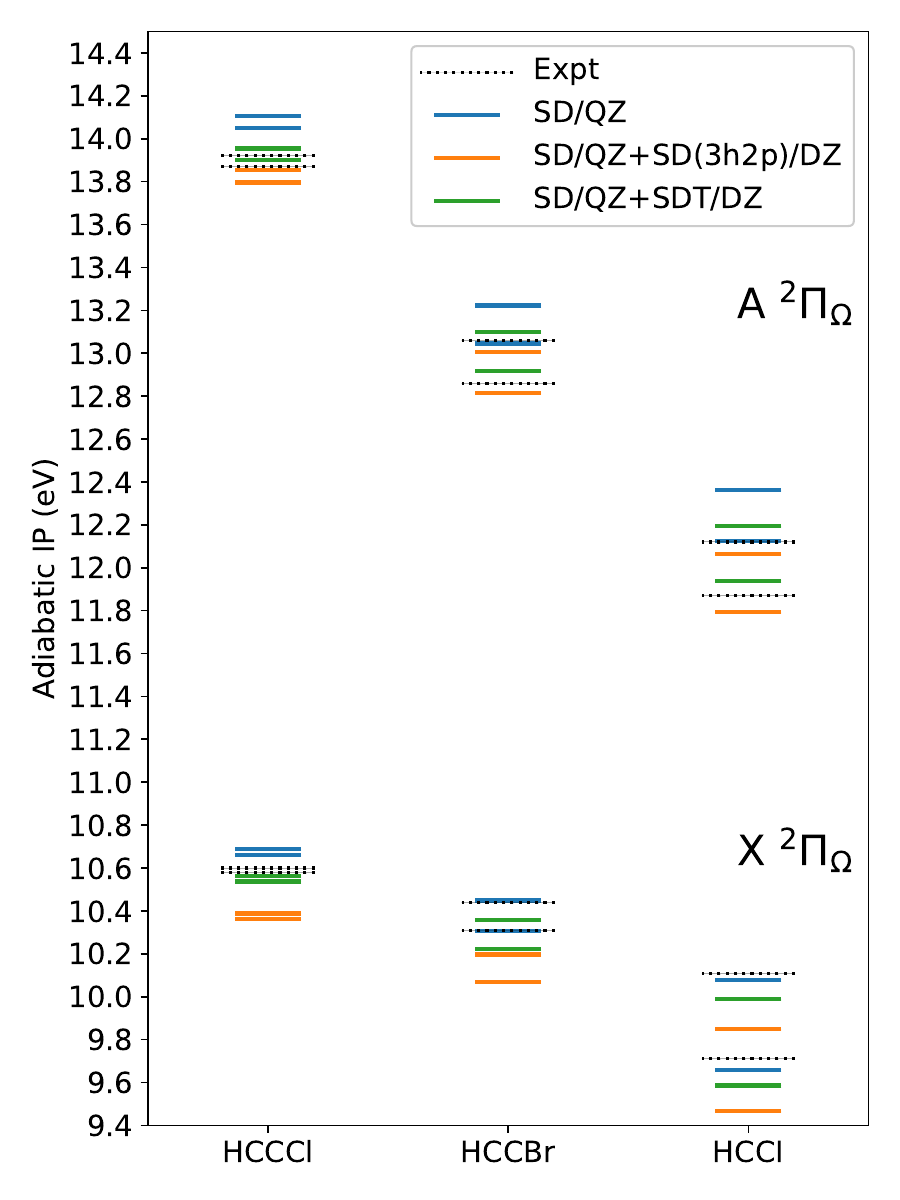}
    \caption{
        \label{fig:composite}
        The calculated adiabatic IPs for HCCCl, HCCBr, and HCCI using DCB-X2C-IP-EOMCCSD/ANO-RCC-VQZP (blue) and the composite schemes described in Eqs.~(\ref{eqn:composite_ip1}) (orange) and (\ref{eqn:composite_ip2}) (green). The experimentally determined IPs from Ref.~\citenum{Maier77_1406} are shown as black dotted lines. The states ordering, from the lowest to highest energy, for each molecule is X $^2\Pi_{\nicefrac{3}{2}}$, X $^2\Pi_{\nicefrac{1}{2}}$, A $^2\Pi_{\nicefrac{3}{2}}$, and A $^2\Pi_{\nicefrac{1}{2}}$.}
\end{figure}

\begin{sidewaystable}[!htbp]
\begin{minipage}{\linewidth}
    \caption{
        \label{tab:composite}
        Adiabatic IPs in eV, spin--orbit splitting ($\Delta_\Omega$) in eV, and 0--0 estimates ($\omega_\mathrm{A-X}$) in cm$^{-1}$ characterizing the HCCX$^{+}$ (X = Cl, Br, and I) cations obtained at selected levels of theory using the DCB-X2C-HF reference. All computed values are reported as errors relative to the available experimental data.}
    \centering
    \begin{tabular*}{\linewidth}{@{\extracolsep{\fill}} llrrrrrrrr}
        \hline\hline
        \multirow{2}{*}{Molecule} & \multirow{2}{*}{Method} & \multicolumn{3}{c}{X $^2\Pi_\Omega$} & \multicolumn{3}{c}{A $^2\Pi_\Omega$} & \multicolumn{2}{c}{A $^2\Pi_\Omega$ -- X $^2\Pi_\Omega$\footnotemark[1]} \\
        \cline{3-5} \cline{6-8} \cline{9-10}
        & & IP $(\Omega = \nicefrac{3}{2})$ & IP $(\Omega = \nicefrac{1}{2})$ & $\Delta_\Omega$
        & IP $(\Omega = \nicefrac{3}{2})$ & IP $(\Omega = \nicefrac{1}{2})$ & $\Delta_\Omega$ 
        & $\omega_\mathrm{A-X} (\Omega = \nicefrac{3}{2})$ & $\omega_\mathrm{A-X} (\Omega = \nicefrac{1}{2})$ \\
        \hline
        \multirow{4}{*}{HCCCl}
        & SD/QZ                          &   0.08  &   0.09  &   0.01  &   0.18  &   0.19  &   0.01  &    810 &   788 \\
        & SD/QZ + SD(3h2p)/DZ            & $-0.22$ & $-0.21$ &   0.01  & $-0.07$ & $-0.07$ &   0.01  &   1162 &  1164 \\
        & SD/QZ + SDT/DZ                 & $-0.04$ & $-0.04$ &   0.01  &   0.03  &   0.03  &   0.01  &    580 &   567 \\
        & Experiment\footnotemark[2]     &  10.58  &  10.60  &   0.02  &  13.87  &  13.92  &   0.05  &  26540 & 26780 \\
        \hline
        \multirow{4}{*}{HCCBr}
        & SD/QZ                          &   0.00  &   0.01  &   0.01  &   0.19  &   0.16  & $-0.02$ &   1516 &  1240 \\
        & SD/QZ + SD(3h2p)/DZ            & $-0.24$ & $-0.24$ &   0.00  & $-0.05$ & $-0.05$ & $-0.01$ &   1552 &  1525 \\
        & SD/QZ + SDT/DZ                 & $-0.09$ & $-0.08$ &   0.00  &   0.06  &   0.04  & $-0.02$ &   1165 &  1003 \\
        & Experiment\footnotemark[2]     &  10.31  &  10.44  &   0.13  &  12.86  &  13.06  &   0.20  &  20570 & 21130 \\
        \hline
        \multirow{4}{*}{HCCI}
        & SD/QZ                          & $-0.05$ & $-0.03$ &   0.02  &   0.26  &   0.24  & $-0.02$ &   2469 &  2199 \\
        & SD/QZ + SD(3h2p)/DZ            & $-0.24$ & $-0.26$ & $-0.02$ & $-0.08$ & $-0.06$ &   0.02  &   1341 &  1658 \\
        & SD/QZ + SDT/DZ                 & $-0.12$ & $-0.12$ &   0.00  &   0.07  &   0.07  &   0.00  &   1555 &  1574 \\
        & Experiment\footnotemark[2]     &   9.71  &  10.11  &   0.40  &  11.87  &  12.12  &   0.25  &  17420 & 16210 \\
        \hline
        \hline\hline
    \end{tabular*}
    \footnotetext[1]{Experimental data correspond to 0--0 transitions. See text for details.}
    \footnotetext[2]{Ref.~\citenum{Maier77_1406}.}
\end{minipage}
\end{sidewaystable}

Adiabatic IP values computed using Eqs.~(\ref{eqn:composite_ip1}) and (\ref{eqn:composite_ip2}) are summarized in Fig.~\ref{fig:composite} and Table \ref{tab:composite}.
These data show that DCB-X2C-IP-EOMCCSD/ANO-RCC-VQZP predicts the adiabatic IPs to the X $^2\Pi_{\nicefrac{3}{2}}$ and X $^2\Pi_{\nicefrac{1}{2}}$ states to within less than 0.1 eV for all the HCCX species investigated here. The good agreement between IP-EOMCCSD and experiment is not too surprising because the lowest ionized state is dominated by the $\ket{\Phi_i}$ determinant where the hole index, $i$, corresponds to the highest-occupied spin orbital. 
However, IP-EOMCCSD overestimates the adiabatic IPs to the A $^2\Pi_\Omega$ states by about 0.16--0.26 eV. The imbalance in the description of the X and A states of the HCCX$^+$ cations translates to $\sim$800, $\sim$1200--1500, and $\sim$2200--2500 cm$^{-1}$ errors in the $\omega_\mathrm{A-X}$ estimates for the chlorine, bromine, and iodine species, respectively, as compared to the experimentally obtained values. Interestingly, the errors in the computed $\omega_\mathrm{A-X}(\Omega = \nicefrac{1}{2})$ values tend to be smaller than those for $\omega_\mathrm{A-X}(\Omega = \nicefrac{3}{2})$. It is also noteworthy that the splitting between the $\Omega = \nicefrac{3}{2}$ and $\Omega = \nicefrac{1}{2}$ states is already well described using DCB-X2C-IP-EOMCCSD/ANO-RCC-VQZP, with errors that are 0.01--0.02 eV in magnitude.

Now, we augment the DCB-X2C-IP-EOMCCSD/ANO-RCC-VQZP IP values using IP-EOMCCSD(3h2p) data computed using the smaller ANO-RCC-VDZP basis set, as defined in Eq.~(\ref{eqn:composite_ip2}). In the IP-EOMCCSD(3h2p) scheme, the neutral HCCX molecules are treated at the CCSD level, while higher-order correlation effects are included in the description of the ionized states. As a result, the X $^2\Pi_\Omega$ states of the HCCX cations are  overstabilized relative to the neutral species; this overstabilization is reflected in the IPs of the X $^2\Pi_{\nicefrac{3}{2}}$ and X $^2\Pi_{\nicefrac{1}{2}}$ states are underestimated by 0.21--0.26 eV (see Table \ref{tab:composite}). On the other hand, the composite scheme reduces the errors in the adiabatic IPs of the A $^2\Pi_\Omega$ states, which are now only 0.05--0.08 eV below the experimentally determined values. Thus, it is not surprising that we do not observe an improvement in the $\omega_\mathrm{A-X}$ values resulting from Eq.~(\ref{eqn:composite_ip2}) when compared to the base DCB-X2C-IP-EOMCCSD/ANO-RCC-VQZP data. Indeed, the errors in $\omega_\mathrm{A-X}$ characterizing the HCCCl$^+$ and HCCBr$^+$ cations actually worsen to $\sim$1200 and $\sim$1500 cm$^{-1}$, respectively, when compared to the $\sim$800 cm$^{-1}$ (HCCCl) and 1200--1500 cm$^{-1}$ errors obtained in the IP-EOMCCSD/ANO-RCC-VQZP calculations. Note that the mismatch in correlation treatments does not negatively impact the $\Delta_\Omega$ splittings, which either improve, in the case of HCCCl and HCCBr, or remain the same, in the case of HCCI.

Correcting IP-EOMCCSD-derived adiabatic IP values with IP-EOMCCSDT instead of IP-EOMCCSD(3h2p) [{\em i.e.}, using Eq.~(\ref{eqn:composite_ip1})] results in a more balanced and accurate description of the X $^2\Pi_\Omega$ and A $^2\Pi_\Omega$ states across all HCCX$^+$ systems examined in this work. The inclusion of both $\hat{T}_3$ and $\hat{R}_{K,\mathrm{3h2p}}$ in IP-EOMCCSDT mitigates the overstabilization of the X $^2\Pi_\Omega$ states that was observed for IP-EOMCCSD(3h2p). Indeed, the magnitudes of the errors in the computed adiabatic IP values for the X $^2\Pi_\Omega$ states are now 0.04, 0.08--0.09, and 0.12 eV for HCCCl, HCCBr, and HCCI, respectively, compared to the >0.2 eV errors resulting from the use of Eq.~(\ref{eqn:composite_ip2}). The A $^2\Pi_{\nicefrac{3}{2}}$ and A $^2\Pi_{\nicefrac{1}{2}}$ states are also well-described, with errors ranging from 0.03 eV for HCCCl to 0.07 eV for HCCI. The improved energies of the individual X $^2\Pi_\Omega$ and A $^2\Pi_\Omega$ translate into better $\omega_\mathrm{A-X}$ estimates relative to experimental data as well. For example, in HCCCl and HCCBr, the errors in the $\omega_\mathrm{A-X}$ values computed using Eq.~(\ref{eqn:composite_ip1}) are only 580 and 567 cm$^{-1}$ in the former case and 1165 and 1003 cm$^{-1}$ in the latter case. These values constitute 2\% and 5\% errors relative to the experimentally observed 0--0 transition energies, which are around 26000 and 21000 cm$^{-1}$, respectively.  For HCCI, IP-EOMCCSDT produces slightly larger errors in $\omega_\mathrm{A-X}$ (1555 cm$^{-1}$), as compared to IP-EOMCCSD(3h2p) (1341 cm$^{-1}$), but both cases offer substantial improvement over the 2469 cm$^{-1}$ error incurred by IP-EOMCCSD/ANO-RCC-VQZP alone. It is also noteworthy that the 1555 and 1574 cm$^{-1}$ errors in $\omega_\mathrm{A-X}$ for HCCI correspond to $\sim$9\% error relative to the experimental $\omega_\mathrm{A-X}$ values. Lastly, we note that the use of IP-EOMCCSDT on top of IP-EOMCCSD either eliminates  error in the spin-orbit splittings ($\Delta_\Omega$ estimates) or leaves the already small errors from IP-EOMCCSD unchanged.

\section{Conclusions}
\label{SEC:CONCLUSIONS}

We have implemented relativistic formulations of IP-EOMCCSD, IP-EOMCCSD(3h2p), and IP-EOMCCSDT where the relativistic treatment is based on the 1eX2C or DCB-X2C formalisms.  
A basis set study on vertical IPs for HCCX [X = Cl, Br, I] molecules at the IP-EOMCCSD level revealed surprising discrepancies between results obtained from the 1eX2C- or DCB-X2C-based calculations when using small contracted basis sets. At the double-$\zeta$ level, 1eX2C- and DCB-X2C-IP-EOMCCSD calculations carried out using contracted x2c- or ANO-RCC-type basis sets produce IP values that differ by up to $\sim$0.1 eV. Calculations using uncontracted Dyall-, x2c-, and ANO-RCC-type basis sets eliminate these differences, which suggests that the error stems from the recontraction of the x2c- and ANO-RCC-style basis sets at the 4c to 2c transformation part of the algorithm. These recontraction issues appear to be less of a problem at larger $\zeta$-levels. Aside from these recontraction issues, we also note that a recent study\cite{Cheng24_054105} has shown the importance of using basis sets that employ spin--orbit contractions instead of the more common spin-free contractions, especially when heavy-element-containing systems are considered. Therefore, the performance of basis sets belonging to the  cc-pV$n$Z-SO\cite{Cheng24_054105} family, which was designed with such contractions in mind, is worth investigating in the future.

We have also shown that the inclusion of correlation effects beyond those captured by IP-EOMCCSD are essential for achieving accurate estimates of IP values in the HCCX series. A DCB-X2C-based composite approach that corrects adiabatic IP-EOMCCSD energetics obtained with a quadruple-$\zeta$ basis set with IP-EOMCCSDT data obtained in a double-$\zeta$ basis set achieves errors relative to experiment on the order of 0.1 eV or less. Remaining error in the computed adiabatic IP values could be attributed to the use of inexact geometries for the neutral and cation species or to the modest size of the basis set used for the IP-EOMCCSDT correction. Large-basis X2C-IP-EOMCCSDT calculations may be challenging to carry out in general, in which case approaches such as the active-space IP-EOMCCSDt approach,\cite{Wloch05_134113,Wloch06_2854,Piecuch06_234107} could be a viable alternative to the composite scheme we have employed. 

Our calculations also demonstrate the effectiveness of the 1eX2C relativistic framework combined with DCB-parameterized SNSO scaling factor in replicating results from the more rigorous DCB-X2C approach; the IP values obtained by both methods are quite similar, especially in the limit of large basis sets. Furthermore, 1eX2C does provide non-negligible savings in memory usage and computational time. For example, in the IP-EOMCCSD calculation for HCCI with Dyall v4z basis set, 1eX2C-IP-EOMCCSD offers a reduction of 21\% in total wall time compared to DCB-X2C-IP-EOMCCSD (see Table S3 in the Supplemental Information), while producing essentially identical IP values and $\Delta_\Omega$ splittings. The computational time savings become 37\%--38\% when the calculations for HCCCl and HCCBr with cc-pV5Z-DK are considered, due to the use of recontracted basis set in the SCF step. Of course, these savings will lessen when considering higher-order correlation effects. Nevertheless, the 1eX2C+SNSO approach appears to be a reasonable alternative to DCB-X2C, particularly when considering larger molecular systems and contracted basis sets.

\vspace{0.5cm}

{\bf Supporting Information} Comparison between IP values obtained from all-electron and frozen-core IP-EOMCCSD/x2c-TZVPPall-2c calculations, IP values obtained using IP-EOMCCSD and the cc-pV$n$Z-DK ($n = {}$D, T, Q, 5) basis sets for HCCCl and HCCBr, and timing information for selected IP-EOMCCSD calculations.

\vspace{0.5cm}

\begin{acknowledgments}This material is based upon work supported by the U.S. Department of Energy, Office of Science, Office of Advanced Scientific Computing Research and Office of Basic Energy Sciences, Scientific Discovery through the Advanced Computing (SciDAC) program under Award No. DE-SC0022263. The Chronus Quantum software infrastructure development is supported by the Office of Advanced Cyberinfrastructure, National Science Foundation (Grant Nos. OAC-2103717 and OAC-2103705). This project used resources of the National Energy Research Scientific Computing Center, a DOE Office of Science User Facility supported by the Office of Science of the U.S. DOE under Contract No. DE-AC02-05CH11231 using NERSC award ERCAP-0027762.\\ 
\end{acknowledgments}

\noindent {\bf DATA AVAILABILITY}\\

    The data that support the findings of this study are available from the corresponding author upon reasonable request.

%\bibliography{bib/Journal_Short_Name,bib/main,bib/deprince,bib/rdm,bib/cc,bib/Li_Group_References}
\bibliography{main}

\end{document}